\documentclass[final,12pt]{elsarticle}
\usepackage{graphicx}
\usepackage{amssymb}
\usepackage{amsmath}
\usepackage{lineno}
\usepackage{xcolor}
\usepackage{subfigure}
 
\journal{Journal Name}
\begin{document}
\begin{frontmatter}
\title{Observation of classically `forbidden' electromagnetic wave propagation and implications for neutrino detection.}

\author[UCI]{S.~W. Barwick}
\address[UCI]{Dept. of Physics \& Astronomy, 4129 Frederick Reines Hall, University of California, Irvine, CA 92697}
\author[UCI]{E.~C. Berg}
\author[KU,MEPHI]{D.~Z. Besson}
\address[KU]{Dept. of Physics and Astronomy, Univ. of Kansas, Lawrence, KS 66045.}
\address[MEPHI]{National Research Nuclear University, Moscow Engineering Physics Institute, 31 Kashirskoye Highway, Rossia 115409}
\author[UCI]{G. Gaswint}
\author[UCI]{C. Glaser}
\author[UPPS]{A. Hallgren}
\address[UPPS]{Uppsala University Department of Physics and Astronomy, Regementsvägen 1, SE-752 37 Uppsala, Sweden}
\author[NIXON]{J.~C. Hanson}
\address[NIXON]{Whittier College Department of Physics, 13406 E. Philadelphia St., Whittier, CA 90602}
\author[LBL]{S.~R. Klein}
\address[LBL]{Lawrence Berkeley National Laboratory,  1 Cyclotron Rd.     Berkeley, CA, 94720}
\author[UCIE]{S. Kleinfelder}
\address[UCIE]{Dept. of Electrical Engineering and Computer Science, University of California, Irvine, CA 92697}
\author[Mainz]{L. K\"opke}
\address[Mainz]{Institute of Physics, University of Mainz, Staudinger Weg 7, D-55099 Mainz, Germany}
\author[UNL]{I. Kravchenko}
\address[UNL]{Dept. of Physics and Astronomy, Univ. of Nebraska-Lincoln, NE, 68588}
\author[Erlangen]{R. Lahmann}
\address[Erlangen]{ECAP, Friedrich-Alexander Universit\"at Erlangen-N\"urnberg, Erwin-Rommel-Str. 1, 91058 Erlangen, Germany}
\author[KU]{U. Latif}
\author[NTU]{J. Nam}
\address[NTU]{Leung Center for Cosmology and Particle Astrophysics; National Taiwan University; No.~1, Sec.~4, Roosevelt Road Taipei, 10617, Taiwan (R.O.C)}
\author[HU,DESY]{A. Nelles}
\address[HU]{Institut f\"ur Physik, Humboldt-Universit\"at zu Berlin, 12489 Berlin, Germany}
\address[DESY]{DESY, Platanenallee 6, 15738 Zeuthen, Germany}
\author[UCI]{C. Persichilli}
\author[UW]{P. Sandstrom}
\address[UW]{Dept. of Physics and Wisconsin IceCube Particle Astrophysics Center, University of Wisconsin, Madison, WI 53706, USA}
\author[UCI,UCIRCC]{J. Tatar}
\address[UCIRCC]{Research Cyberinfrastructure Center, University of California, Irvine, CA 92697}
\author[UPPS]{E. Unger}

\begin{abstract}
Ongoing experimental efforts in Antarctica seek to detect ultra-high energy neutrinos by measurement of radio-frequency (RF) Askaryan radiation generated by the collision of a neutrino with an ice molecule. An array of RF antennas, deployed either in-ice or in-air, is used to infer the properties of the neutrino. To evaluate their experimental sensitivity, such experiments require a refractive index model for ray tracing radio-wave trajectories from a putative in-ice neutrino interaction point to the receiving antennas; this gives the degree of signal absorption or ray bending from source to receiver. 

The gradient in the density profile over the upper 200 meters of Antarctic ice, coupled with Fermat's least-time principle, implies ray ``bending'' and the existence of ``forbidden'' zones for predominantly horizontal signal propagation at shallow depths. After re-deriving the formulas describing such shadowing, we report on experimental results that, somewhat unexpectedly, demonstrate the existence of electromagnetic wave transport modes from nominally shadowed regions.
The fact that this shadow-signal propagation is observed both at South Pole and the Ross Ice Shelf in Antarctica suggests that the effect may be a generic property of polar ice, with potentially important implications for experiments seeking to detect neutrinos.
\end{abstract}

\end{frontmatter}

\section{Introduction}
Owing to its remote location and isolation from anthropogenic sources, excellent transparency at wavelengths ranging from optical through radio, and also the presence of extensive scientific support at several locations, Antarctica now supports multiple astronomy and astrophysics-oriented projects.
Within the last five years, the IceCube experiment, sensitive to optical and near-optical
Cherenkov radiation resulting from neutrino interactions in-ice, has reported on the first observation of a diffuse flux of extraterrestrial neutrinos at energies greater than 10 TeV \citep{2016ApJ...833....3A}, with a `hard' spectrum extending to $10^{15}$ eV.  At higher energies, in-ice detection of longer-wavelength (radio) radiation is likely a more sensitive measurement strategy, owing to the Askaryan effect \citep{Askaryan1962a,Askaryan1962b,Askaryan1965}, combined with the measured kilometer-scale radio-wave attenuation length for cold polar ice \citep{barrella2011ross,barwick2005south}. 
This has prompted several experimental initiatives based on experimental radio receiver arrays either elevated 35--40 km  (ANITA \citep{GorhamAllisonBarwick2009}), near the Antarctic ice-air interface at Moore's Bay, Antarctica (ARIANNA \citep{Barwick:2014pca}), or at depths of up to 200 m at South Pole (pioneering RICE \citep{KravchenkoFrichterSeckel2003} and successor ARA \citep{allison2016performance:2015eky}). 
In addition, exploratory work has been conducted within the last few years at Summit, Greenland to assess the radio-glaciological suitability of that site for a future neutrino-detection experiment \citep{avva2015situ}.
Each of the possible neutrino-observation schemes (synoptic, surface detection of antennas, or antennas embedded in the ice sheet) has its own inherent advantages and trade-offs. 

\begin{figure}
\centering
\begin{subfigure}[]{\includegraphics[width=0.5\textwidth]{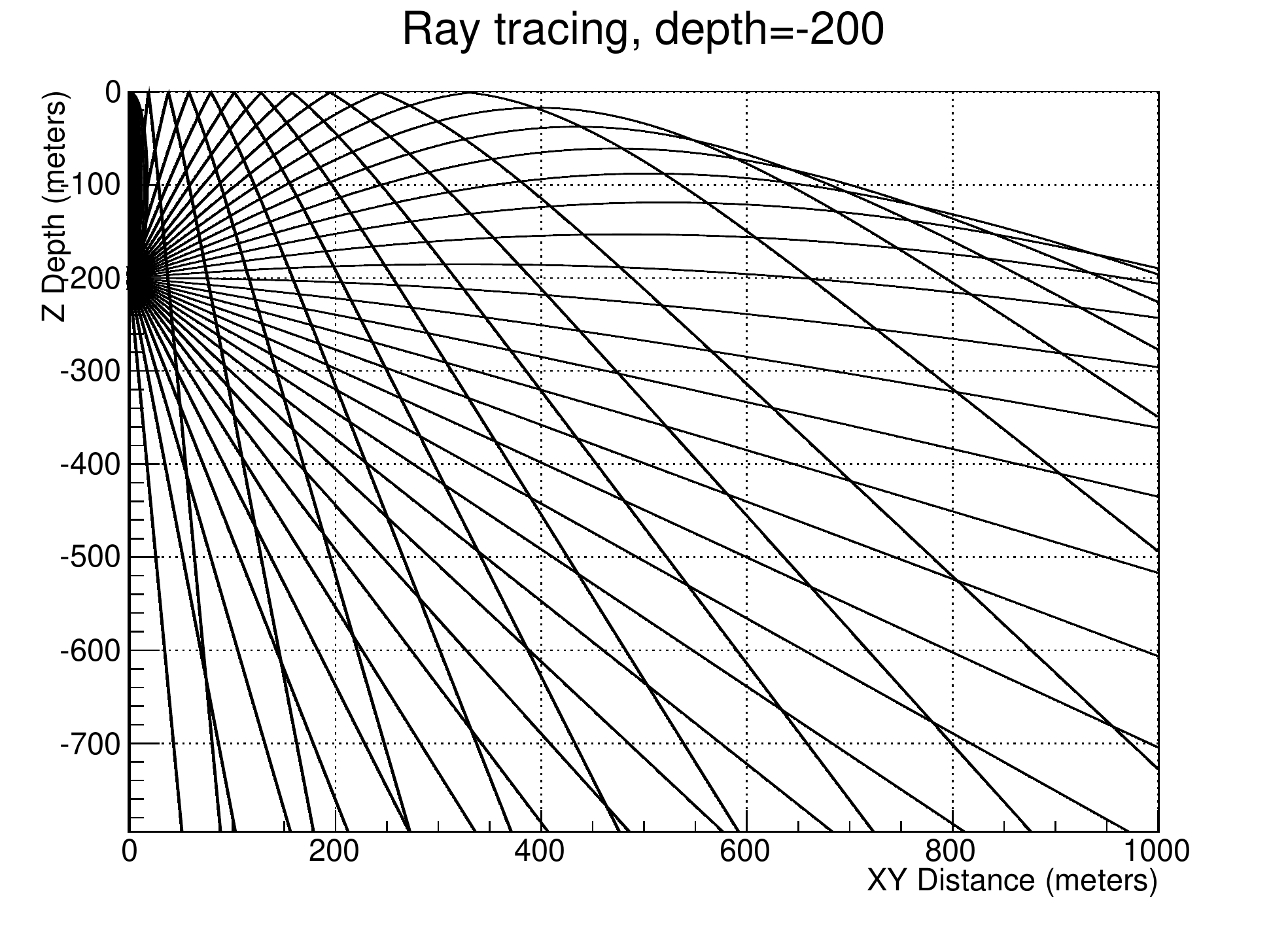}}
\end{subfigure}
\begin{subfigure}[]{\includegraphics[width=0.37\textwidth,angle=270]{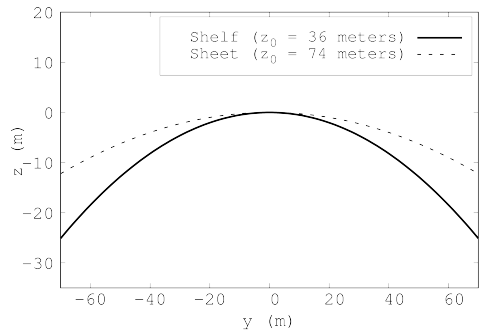}}
\end{subfigure}
\caption{\label{fig:RayTracing} (a) Simulation of rays emitted from a transmitter at the South Pole at z=-200 m, showing curved paths. The shadow zone in the upper right is expected in the case of a smoothly-varying $n(z)$ profile in the absence of impurities. For the RICE measurements described later, the XY-distance was approximately 3000 meters, at typical depths of 100 meters (b) Examples of quadratic ray-paths in media with index of refraction profiles with the form of Eq.~\ref{eq:n}.  The dashed line corresponds to a particular solution of Eq.~\ref{eq:ray_tr} with $z_{\rm 0} = 74$ m.  The solid line corresponds to Eq.~\ref{eq:ray_tr} with $z_{\rm 0} = 36$ m.}
\end{figure}

The variable specific gravity through the firn \citep{pearce1967empirical}, over which the ice density varies between approximately 40--100\% of the asymptotic value (917 $\mathrm{kg}/\mathrm{m}^3$), results in an electromagnetic wave-speed decreasing with depth. By Fermat's principle, this results in not only curved ray trajectories, but also the expectation that, for the case where transmitter (Tx) and/or receiver (Rx) is deployed either on the surface or at near-surface depths, signals emanating from sufficiently large horizontal angles may be refracted downwards before they can be observed (``shadowing''). In the Huygens picture, these shadowed regions correspond to volumes for which the superposition of all contributing wavelets, properly weighted by distance, sum to zero net amplitude for all observation times $t$, as illustrated in Fig.~\ref{fig:RayTracing}.

We note that the standard Huygens picture is typically applicable in the case where phase information is preserved by each scatterer, assumed to be small compared to one wavelength (i.e., the Rayleigh limit), and results (assuming zero signal absorption in the medium) in the usual $1/r$ length scaling of the electric field ${\vec E}$. If the scatterer is not point-like (e.g., scattering size $\sim\lambda$) or, if, for any other reason, the phase shift across the scatterer is random, this leads to $|{\vec E}|\propto 1/r^2$. 

For dielectric materials like snow and ice, the signal wave-speed is determined by the local index of refraction, which can be approximated as a linear equation of density: $n(z) \approx 1+b\rho(z)$, with $z$=0 at the surface and increasingly negative with depth.  The specific dependence for ice is given by the Schytt equation: $n(x,y,z)=1+0.78\rho(x,y,z)/\rho_0$, where $\rho(x,y,z)$ is the local ice density and $\rho_0$ is the density for solid ice (917 $\mathrm{kg}/\mathrm{m}^3$).  Designating $n_s$ as the index of refraction of snow (see Sec.~\ref{section:icedensity}), $n_{ice}$ that of solid ice, and $\Delta n = n_{ice} - n_{s}$, then it can be shown from classical gravity and density considerations that the index versus scale depth $z/z_0$ ($z_0 > 0$) dependence follows:
\begin{equation}
n(z) = n_{ice} - \Delta n e^{z/z_0}~.
\label{eq:n}
\end{equation}

From the same classical treatment that produces Eq.~\ref{eq:n}, it may be shown that $z_{0}^{-1} = (g\chi_0\rho_s)$, where $g$ is the gravitational acceleration, and $\rho_s$ and $\chi_0$ are the density and volumetric compressibility of snow, respectively.  The snow density and compressibility are inversely proportional, while measurements of natural snow compressibility vary in the literature and depend on the measurement technique \citep{mellor1974review}.  Taking compressibility values from fit F of Fig.~3 of \citep{doi:10.1002/2017GL075110} at $\rho_s = 300$ kg m$^{-3}$ yields a $z_0$ value of 25 m. Rather than measure $\chi$ and $\rho$ independently, we fit $z_0$ as a free-parameter obtained from $\rho(z)$ data from various locations around Antarctica (see Fig.~\ref{fig:nvsz} and Tab.~\ref{tab:tab0}).  We find agreement with prior measurements \citep{barrella2011ross,hansJGlac}, and also find that $z_0$ varies by a factor of $\approx$ 2 between Moore's Bay and the South Pole.  Snow formation conditions near the surface vary considerably across polar regions, so there is no reason to expect the compressibility of surface snow to be uniform across different glaciological regions.

Allowing $\chi$ to vary with $z$ in the density versus depth model yields the following boundary-value relation for $\chi_s$, the compressibility of surface snow, $\chi_{ice}$ the compressibility of deep ice, and $\chi_{f}$, the compressibility of the firn:
\begin{equation}
(\Delta\rho)\chi_f = \rho_s \chi_s - \rho_{ice}\chi_{ice}~.
\label{eq:chi}
\end{equation}
In Eq.~\ref{eq:chi}, $\chi_f$ is a density-weighted difference between snow and ice compressibility, which serves as a useful average for the firn, as a whole.  Although the depth-dependence of the compressibility of the firn $\chi(z)$ is outside the scope of this work, we note that if $\chi$ depends monotonically on depth, Eq.~\ref{eq:n} disallows horizontal ray tracing solutions.  To explain horizontal ray tracing, a perturbation in the index profile can be added to Eq.~\ref{eq:main} (see Sec.~\ref{sec:F}).

\section{Formalism}
\label{sec:F}
In this section, ray tracing theory is reviewed.  We begin with Fermat's principle and conclude with a discussion of conditions that lead to horizontal ray propagation, anticipating the experimental results described below.

\subsection{Fermat's Principle and ray tracing}
Fermat's Principle states that optical lengths of light ray trajectories are minimized.  Ray paths that satisfy Fermat's Principle depend on the index of refraction $n$.  If $n$ depends only on $z$, Fermat's Principle can be expressed in variational form as:
 
\begin{align}
\delta &\int_A^B n(z)(1+\dot{y}^2)^{1/2} dz = \delta \int_A^B L(z,\dot{y}) dz = 0~.
\end{align}

Derivatives indicated with a dot are with respect to $z$. Because $n(z)$ does not depend on $x$ or $y$, the problem exhibits cylindrical symmetry.  Without loss of generality we can choose $x = \dot{x} = 0$.  Note that $\dot{y} = dy/dz$ is unit-less, and $\ddot{y}$ has units of inverse length.  Minimizing the variation in the path, and substituting $u = \dot{y}$ gives
\begin{equation}
\dot{u} = - \left( \frac{\dot{n}}{n} \right) (u^3+u). \label{eq:main}
\end{equation}

Inserting Eq.~\ref{eq:n} for $n(z)$, the equation of motion is

\begin{equation}
\dot{u} = z_0^{-1} \left( \frac{\Delta n e^{z/z_0}}{n_{ice} - \Delta n e^{z/z_0}} \right) (u^3+u)~. \label{eq:dotu}
\end{equation}

As a check, note the deep ice limit: $|z| \gg z_0$, $z<0$:

\begin{equation}
\dot{u} = 0~.
\end{equation}

The solution to this equation of motion is 

\begin{equation}
z(y) = a+by~.
\label{eq:straight}
\end{equation}

Eq. \ref{eq:straight} shows that rays propagate in straight lines far below the firn where $n$ is constant, as expected \footnote{Note that a horizontal solution to Eq.~\ref{eq:dotu} would imply $\dot{u} \to \infty$, requiring $\dot{n} \to 0$ in Eq.~\ref{eq:main}.  However, $\dot{n} = 0$ cannot occur without an under-density or over-density in the firn, since index and density are proportional.}.  Another straight-line solution to Eq.~\ref{eq:dotu} is the vertical ray ($u = 0$), which remains straight while progressing through all regions of $n(z)$. 

For the case of a shallow ray ($z \rightarrow 0$) with $n \approx 
n_{ice} - \Delta n (1+ z/z_0)$ and $\dot{n} \approx - \Delta n /z_0$
initially propagating with a horizontal velocity component satisfying $u^3 \gg u$, the main equation of motion (Eq.~\ref{eq:dotu}) reduces to 

\begin{equation}
\frac{du}{dz} = \frac{1}{z_0}\left( \frac{\Delta n}{n_{ice} - \Delta n (1+ z/z_0)}\right) u^3~,
\end{equation}

Keeping only first order terms in $(z/z_0)$, a particular solution is

\begin{equation}
z(y) = -\frac{1}{2z_0} \left( \frac{n_{ice} - n_s}{n_s}\right) (y-y_1)^2~.
\label{eq:ray_tr}
\end{equation}

Equation \ref{eq:ray_tr} shows that the shortest travel time between two near-surface points is given by a quadratic path, if the initial velocity vector is mostly horizontal.

For example, take $z_0 = 36$ m and $n_s = 1.30$ to describe refraction at Moore's Bay, Antarctica (site of the ARIANNA experiment), and $z_0 = 74$ m, $n_s = 1.35$ to describe South Polar refraction (see Tab.~\ref{tab:tab0} for measured values).  These two ray paths are compared in Fig.~\ref{fig:RayTracing}.  The curvature of the quadratic in Eq. \ref{eq:ray_tr} is controlled by $z_0^{-1}$.

The ray tracing framework yields near-surface ray paths that are downward bending quadratic curves, with \textit{smaller $z_0$ values corresponding to steeper bending}.  The data presented in Sec. \ref{sec:observations} include observations of rays that not only do not propagate with quadratic downward bending, but propagate horizontally in Moore's Bay where the value of $z_0$ is approximately a factor two smaller than that of the South Pole.  If rays are not shadowed in Moore's Bay, it should be even less likely that they are shadowed in the firn of the South Pole, and data presented in Sec. \ref{sec:observations} also support this hypothesis.

\subsection{Horizontal and Near-Surface Propagation}

Perturbations from the smooth profile can be introduced by variable yearly melting and sintering mechanisms, and bulk re-alignment of the crystal orientation fabric. Chapter 2 of \citep{Bogo85} summarizes these mechanisms, and such observations of layers are common \citep{gerland_oerter_kipfstuhl_wilhelms_miller_miners_1999, 0034-4885-67-10-R03}.  We observe layering in Moore's Bay and South Pole data as $\approx 5$ \% deviations from a smooth fit to the density profile (see Figs.~\ref{fig:nvsz} and \ref{indexofrefr_1}). Over-densities (such as those observed near the South Polar surface) and under-densities can lead to local minima and maxima in the index of refraction profile. 

Let one such local feature be described by a quadratic perturbation from an otherwise constant $n_{\rm 0}$ value, with a strength $a$ at a depth $z_{d}$:

\begin{align}
n(z) &= n_0 + a(z-z_{d})^2 \label{eq:perturb} \\
\dot{n} &= 2a(z-z_{d}) \\
q &= z-z_{d}~,
\end{align}

Let $\omega^2 = 2\left(\frac{a}{n_0}\right)$.  Introducing $n(z)$ from Eq.~\ref{eq:perturb} into Eq.~\ref{eq:main}, and neglecting terms higher than order $q^2$, the variables-separable differential equation may be solved near $q=0$:

\begin{equation}
q(y) = C_0 \sin(\pm C_1\omega y - C_2)
\label{eq:SHO}
\end{equation}

The constants $C_i$ are determined by the boundary conditions and the shape of the perturbation, and two of them are independent.  The approximation is accurate as long as $\omega^4 q^4 \ll 1$.  Solving the problem in the same limit with $a \rightarrow -a$ in Eq.~\ref{eq:perturb} amounts to replacing the sine function with a sinh function in Eq.~\ref{eq:SHO}, making the path $q(y)$ unbounded.

A quadratic perturbation in $q$ can only be added in a piecewise-continuous fashion to Eq.~\ref{eq:n}, if the boundary conditions $n \rightarrow n_{ice}$ as $z \rightarrow -\infty$ and $n(0) = n_s$ are to be preserved.  Admitting a Gaussian perturbation yields the physical behavior of the locally quadratic perturbation, while keeping $n(z)$ fully continuous and differentiable:

\begin{equation}
n(z) = n_0 + a \exp\left(-\frac{1}{2}\left(\frac{q}{\sigma}\right)^2\right) \label{eq:perturb2}
\end{equation}

The prior definition of $\omega$ with the $a$-value from Eq.~\ref{eq:perturb} has units of inverse length.  The $a$-value in Eq.~\ref{eq:perturb2} is unit-less, but the perturbation width $\sigma$ has units of length.  Repeating the procedure leading to Eq.~\ref{eq:SHO}, in the limit that $(\omega q/\sigma)^4 \ll 1$, the solution is 

\begin{equation}
q(y) = C_0 \sin\left(\pm C_1\frac{\omega y}{\sigma} - C_2\right)
\label{eq:SHO2}
\end{equation}

If horizontal ray-propagation were observed, there would be several potential conclusions. First, the $n(z)$ profile could be described by Eq.~\ref{eq:n} with local density perturbations.  According to Eq.~\ref{eq:SHO2}, the rays would oscillate about the perturbation with a spatial frequency and amplitude determined by the shape of the perturbation.  A second possibility is that the perturbations could have such large $a$-values and such small $\sigma$-values that rays are simply reflected by them.  Groups of such internal layers could form reflective channels, trapping rays in horizontal states through total internal reflection.  Although we do not discern from the data which mechanism is present in the ice sheets and ice shelves, we note that ice layers are common in the upper firn, \textit{and} that over- and under-densities do appear in residual fits of Eq.~\ref{eq:n} to the $n(z)$ data (see Fig.~\ref{fig:nvsz}).

\subsection{Density and propagation-time measurements in Antarctica}
\label{section:icedensity}

\begin{figure}
\centering
\includegraphics[width=0.7\textwidth]{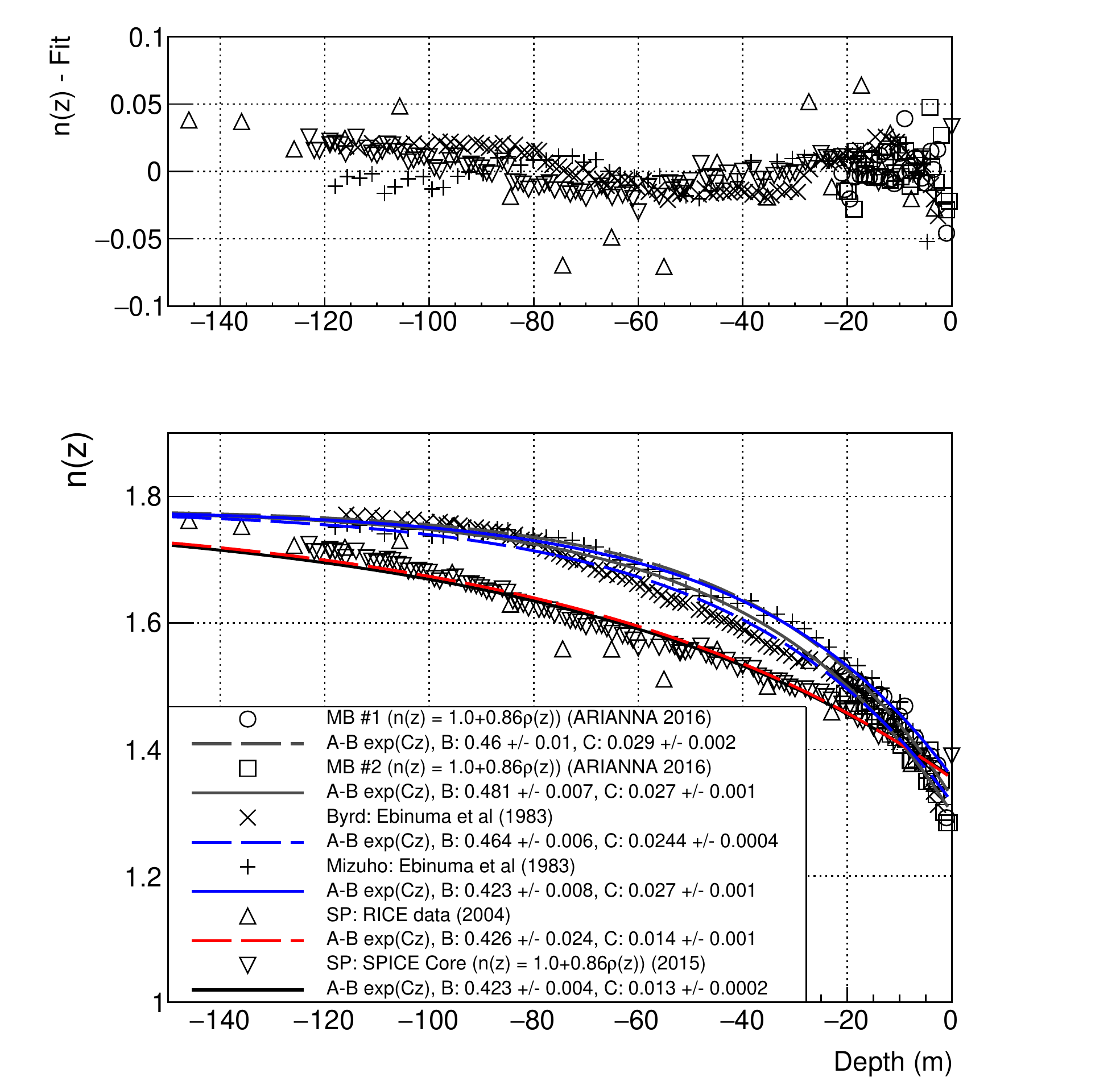}
\caption{\label{fig:nvsz} Compilation of density and index of refraction measurements.  ``MB'' results (circles and squares) refer to density measurements made by the authors during the 2016-17 austral season at Moore's Bay, Antarctica, and expressed here as index of refraction via the Schytt equation.  The Byrd and Mizuho density data (x's and crosses) are taken from \citep{maeno1983pressure} and translated to index of refraction in the same fashion.  The RICE data (triangles) are direct measurements of index using RF signals from \citep{kravchenko2004situ}.  The SPICE-core data (upside-down triangles) come from the 2015 SPICE core density measurements from the South Pole, and are translated via the Schytt equation. The residual difference between the fit lines and the data are plotted in the upper panel.}
\end{figure}

Measurements of density and index of refraction have been compiled in Fig.~\ref{fig:nvsz} for a variety of Antarctic locations. Table \ref{tab:tab0} contains the coefficients $A=n_{ice}$, $B = \delta n$, and $C = z_0^{-1}$ determined from a fit of the form $n(z) = A-B\exp(Cz)$.  Relative to a smooth functional dependence, variations in measured density are observed at the level of a few percent, 
larger than the intrinsic systematic errors (estimated at less than 1\% relative), and decreasing with depth.

\begin{table}[ht]
\begin{center}
\begin{tabular}{| c | c | c | c |}\hline
Ref./Location & $B = \Delta n$ & $n_s = n_{ice} - B$ & $C^{-1} = z_0$ (m)\\ \hline
MB\#1/Moore's Bay & $0.46\pm0.01$ & $1.32\pm0.01$ & $34.5\pm2$ \\ \hline
MB\#2/Moore's Bay & $0.481\pm0.007$ & $1.299\pm0.007$ & $37\pm1$ \\ \hline
Ebinuma (1983)/Byrd & $0.464\pm0.006$ & $1.316\pm0.006$ & $41\pm1$ \\ \hline
Ebinuma (1983)/Mizuho & $0.423\pm0.008$ & $1.357\pm0.008$ & $37\pm1$ \\ \hline\hline
RICE (2004)/South Pole & $0.43\pm0.02$ & $1.35\pm0.02$ & $71\pm5$ \\ \hline
SPICE (2015)/South Pole & $0.423\pm0.004$ & $1.357\pm0.004$ & $77\pm2$ \\ \hline
\end{tabular}
\caption{\label{tab:tab0} Fit parameters for the curves shown in Fig.~\ref{fig:nvsz}.  The function fit to the data is $n(z) = A-B\exp(Cz)$.  The second-order differential equation derived in the first section requires A=$n_{ice} = 1.78$ and $B = \Delta n$ as the two boundary conditions.}
\end{center}
\end{table}

In Moore's Bay, the parameter $n_s$ (index of refraction of the snow near the surface), has been measured in two ways.  First, surface snow density measurements were recorded and converted to index via the usual Schytt equation to determine $n_s = 1.3$ \citep{Gerhardt:2010js}.  Second, the absolute timing of an RF pulse transmitted at 2--4 wavelengths below the surface through the snow along a 543 meter baseline corresponded to a measurement of $n_s = 1.29\pm 0.02$ \citep{hansJGlac}.  These results are in agreement with the fits to the density data versus depth shown in Fig.~\ref{fig:nvsz} and Tab.~\ref{tab:tab0}, and in agreement with $n_s$ values obtained from density measurements at the South Pole and two other locations.

\begin{figure}
\includegraphics[width=0.9\textwidth]{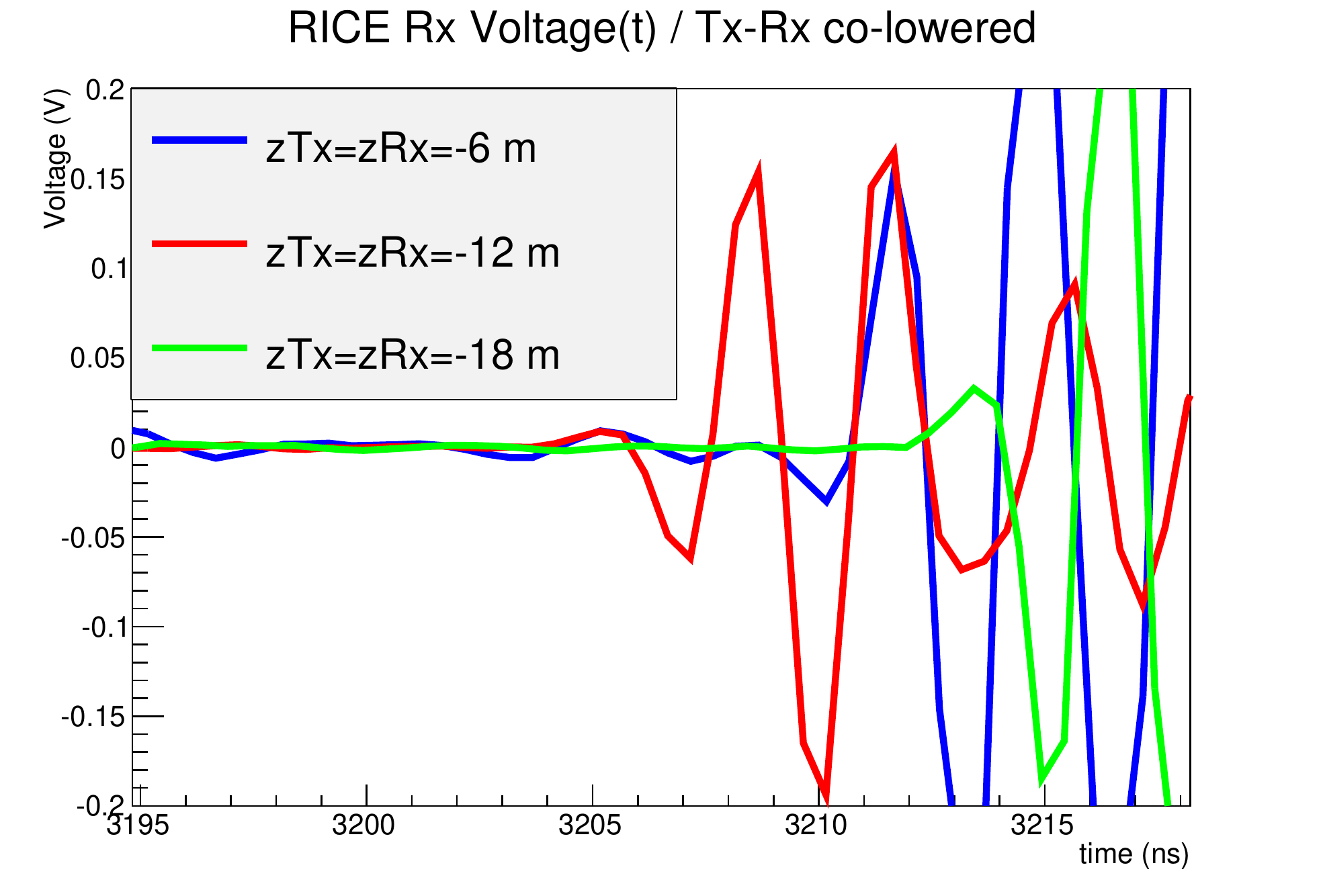} 
\caption{Received signals for cases where transmitter/receiver depths are 6 m (red), 12 m (blue) and 18 m (green), from dedicated, 2003 data taken with RICE experimental hardware, illustrating earlier arrival time for 12-meter depth compared to 6-meter depth, consistent with observed local fluctuation in SPICE density profile. Horizontal separation between transmitter and receiver is approximately 30 m for these data. Systematic error on relative signal arrival time is of order 0.1 ns.}
\label{fig:B2XB4}
\end{figure}

The RICE data presented in Fig.~\ref{fig:nvsz} was collected with a ~0.5-km distance between RF transmitter and receiver, and relied on relative timing between stationary RF receiver channels as a single transmitter is lowered into an ice borehole.  A direct RF-based measurement of the South Pole $n(z)$ was conducted in December 2003 using two RF antennas co-lowered into boreholes separated by 30 m horizontally; those data are presented in Fig.~\ref{fig:B2XB4}.  
The absolute timing between RF transmitter and receiver provides a direct measurement of $n(z)$. By contrast, the ``MB'' data in Fig.~\ref{fig:nvsz} is density data that has been converted to $n(z)$ via the Schytt equation.  The near-surface comparison of the 2003 RICE measurements with the density data is particularly interesting -- here, the local-minimum in SPICE density measurements at z=--12 m suggestively correlates with a local minimum in RF propagation time at that same depth. 
This is consistent with the presence of `inversion' layers which, as demonstrated in Section 2 of this document, owing to the least-time principle, can result in signals arriving horizontally from nominally `shadowed' regions.


\section{Observation of signal propagation from shadow zones}
\label{sec:observations}
Experimental measurements of the radio-frequency dielectric permittivity have been made over the last 15 years in Antarctica \citep{barwick2005south} and also more recently in Greenland \citep{avva2015situ}. In those previous measurements, radio wave signals propagate vertically from a surface or near-surface transmitter, and are observed in a surface or a near-surface receiver via their reflection either from an in-ice horizontal conducting layer, e.g., or the underlying bedrock. This approach has the advantage that the transmitter and receiver can be easily moved on the surface, and flexible triggers configured. However, since the solid angle for neutrino acceptance varies with polar angle $\theta$ as $\sin\theta$, the neutrino effective target volume becomes diminishingly small viewing vertically, and such measurements therefore have limited applicability to neutrino sensitivity estimates.

\subsection{Measurements made by the RICE experiment at South Pole}

Given its importance vis-a-vis neutrino sensitivity, verification of shadowing was given high priority in the early stages of the RICE experiment \citep{kravchenko2004situ,KravchenkoFrichterSeckel2003}.
During the period Dec.~2003 -- Jan.~2004, microsecond-duration ``tone'' signals were transmitted horizontally over a baseline of 3.3 -- 3.5 km, at depths of 70, 120 and 125 meters from a borehole drilled originally for the National Oceanic and Atmospheric Administration (NOAA). 
The 20-channel RICE antenna array, based at South Pole and including 17 receiver antennas deployed at depths between 105 and 350 meters, was located in the nominal 'shadow' zone, as evident from Fig.~\ref{fig:RayTracing} above.

For reference, and to simplify a calculation of attenuation length based on relative received signal strengths, data were also collected, using exactly the same transmitter set-up, from a closer location embedded within the RICE array itself, and unshadowed. With two such transmitter locations, $L_{atten}$ can be numerically extracted using the ratio of signal amplitudes measured at the far transmitter location ($A_{\mathrm{far}}$) relative 
to the `near' transmitter location ($A_{\mathrm{near}}$), and assuming that electric field strengths vary inversely with distance \begin{equation}
A_{\mathrm{far}}/A_{\mathrm{near}}=(\cos\theta_{Tx\to Rx}^{\mathrm{near}}/\cos\theta_{Tx\to Rx}^{\mathrm{far}})|r_{\mathrm{near}}/r_{\mathrm{far}}|\times e^{-(r_{\mathrm{far}}-r_{\mathrm{near}})/L_{\mathrm{atten}}},
\end{equation}
with the values of $r$ defined individually for each Tx/Rx pair; the $\cos\theta$ term accounts for the antenna dipole beam pattern of the dipoles. 


Broadcast signals were produced as follows: 
\begin{enumerate}
\item A signal generator (\emph{SG}), producing continuous waves in the interval 211$\to$500 MHz, is gated open once per second by a GPS pulse per second (pps) trigger, for a period of between one and 20 microseconds. 
\item This signal generator output is then split into two copies: one copy is routed to an above-surface TV log-periodic-dipole-antenna (LPDA) (\emph{TV}) pointed at a similar above-surface receiver LPDA antenna co-located with the RICE receiver array and fed (arbitrarily) into RICE channel 11. Receipt of that above-ice signal in channel 11 provides the event trigger for the RICE array, initiating readout of the remaining channels.
\item The second copy is passed through a 100W amplifier, and then routed into a 300-meter length of 7/8" Andrews coaxial cable, at the end of which is the buried RICE Dipole (\emph{DI}) antenna transmitter, efficient over the interval 200-500 MHz, and used to broadcast under-ice signal to the RICE Dipole (\emph{DI}) receiver array.
An additional delay unit staggers the $SG\to TV$ vs.~$SG\to DI$ signals to ensure that they are emitted roughly simultaneously. 
\end{enumerate}

In principle, multiple signal paths are possible from the two transmitters to the RICE receivers, which we designate as $TV\to TV$ (signals measured in the above-air
receiver from the above-air TV transmitter, and providing the RICE event trigger), 
$TV \to DI$ (signals measured in the in-ice RICE Dipole receiver channels from the above-air TV transmitter), and $DI\to DI$ (signals measured in the in-ice RICE Dipole receiver channels from the in-ice Dipole transmitter).
These multiple signal paths are indeed seen as
signals in the RICE channels.

\begin{figure}\centerline{\includegraphics[width=\textwidth]{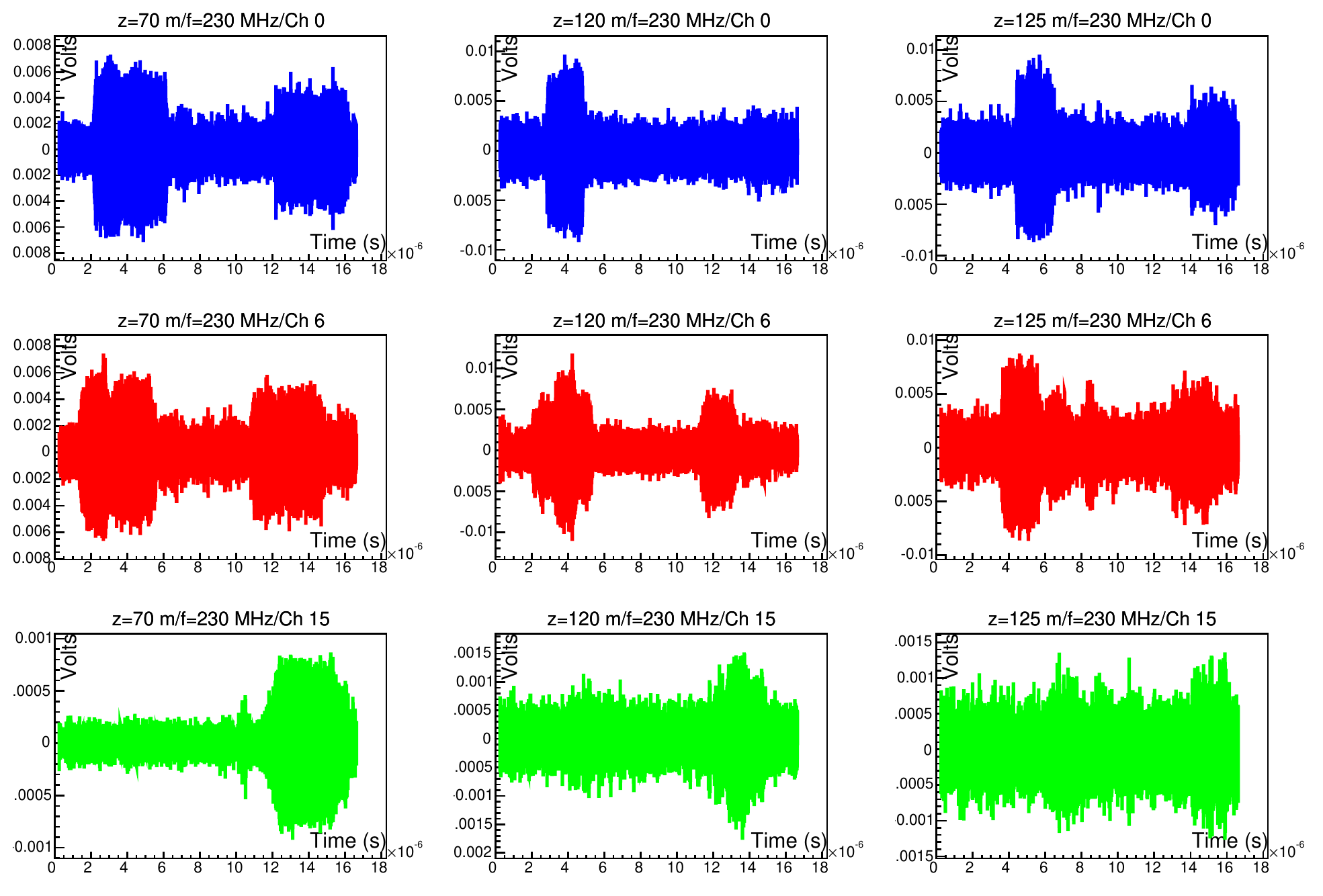}}
\caption{Phase-aligned sum of signals observed for three RICE channels, with transmitter at indicated depths $-70$ m, $-120$ m, or $-125$ m. Receiver depths are $-166$ m (Ch 0), $-170$ m (Ch 6) and $-367$ m (Ch 15), respectively.}
\label{fig:XSpreso_Avsz}
\end{figure}

\begin{figure}\centerline{\includegraphics[width=\textwidth]{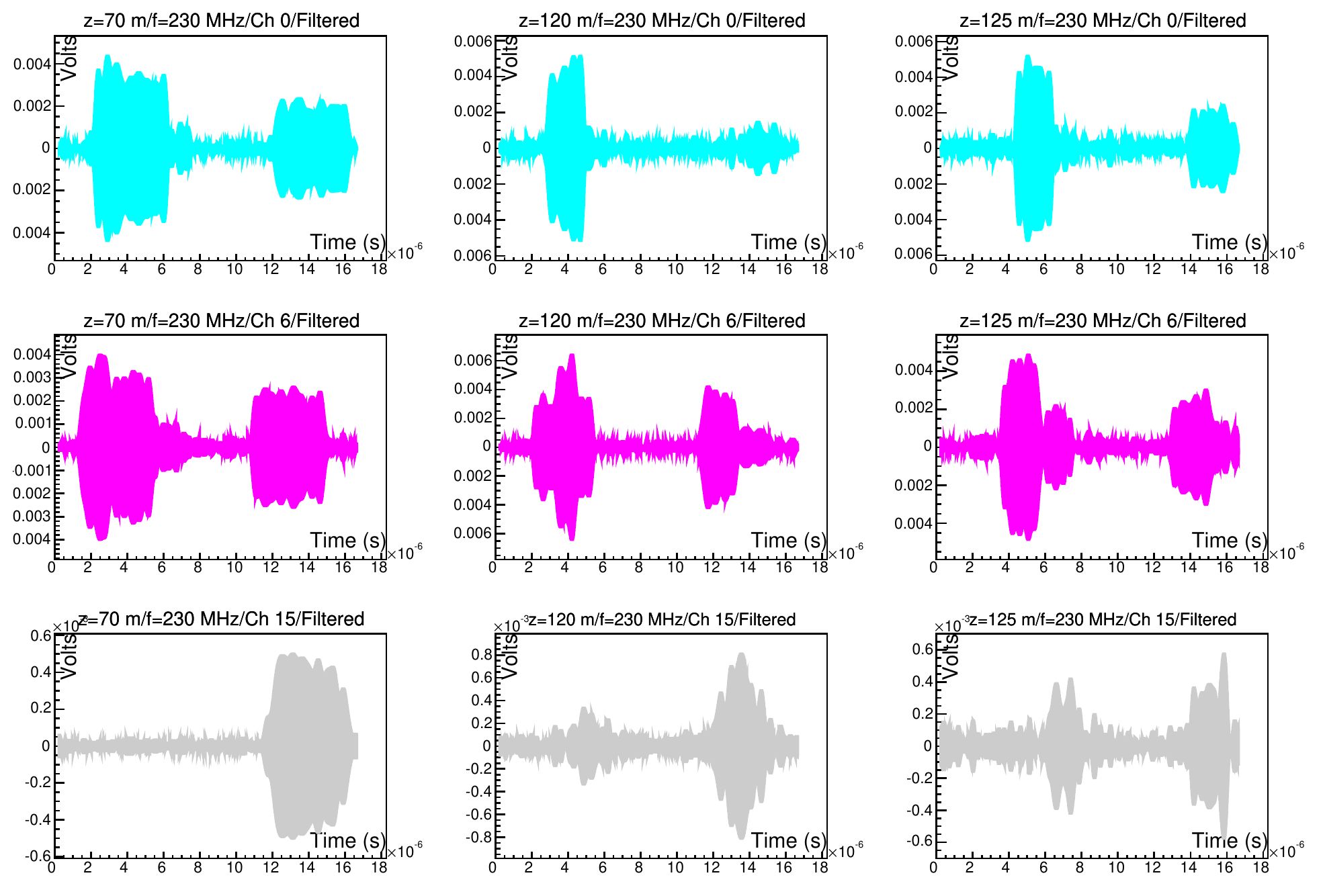}}
\caption{Same signals as in previous plot, after filtering $\pm$2 MHz around known broadcast frequency. Vertical axis: Volts; Horizontal axis: time (s)}
\label{fig:XSpreso_AvszF}
\end{figure}

Although signals cannot be averaged during data-taking, to improve the signal-to-noise of the $DI\to DI$ signal, the in-ice receiver traces were phase-aligned, event-by-event, using the event-by-event relative phase shifts derived from the bright $TV \to TV$ signals, 
which are observed as nearly pure sinusoids. Fig.~\ref{fig:XSpreso_Avsz} shows the signals observed in three RICE channels (channel 0: top row, receiver $z=-166$ m; channel 6: middle row, receiver $z=-170$ m; channel 15: bottom row, receiver $z=-367$ m), for data taken at the three transmitter depths (left column: $z=-70$ m,  middle column: $z=-120$ m, and right column: $z=-125$ m), after phase alignment. Fig.~\ref{fig:XSpreso_AvszF} shows the same traces after filtering around the carrier.


Clearly observed in these dipole receiver plots are two signals, offset in time by approximately 10 microseconds, which we interpret as the $TV \to DI$ and $DI\to DI$ signals, respectively. 

\subsubsection{Comparison with Expectation}
Since the entire RICE array is in the nominal shadow zone for this geometry, observation of the latter of these signals is immediately in conflict with naive expectations. For non-shadowed signal propagation, received signals should: i) be of time duration identical to the broadcast signal, and ii) for each RICE channel, exhibit a signal strength which is independent of the depth of the transmitter, since the difference in path-length between 70 meter transmitter depth (the experimental minimum) and 125 meter transmitter depth (the experimental maximum) is negligible, given the approximately 3 km horizontal propagation baseline. By contrast,
the experimentally received signals show unexpectedly large variation in amplitude, compared to the simple 1/r expectation, for relatively small vertical displacements (120 m vs.~125 m, e.g.); a model that adequately describes these variations is currently under development. We note that the large variations observed in received signal strength, for 5-meter variations in transmitter depth, cannot be explained by channel-to-channel gain uncertainties, as those uncertainties are inherent, and identical for the transmitter at any depth.

\begin{figure}
\includegraphics[width=0.48\textwidth]{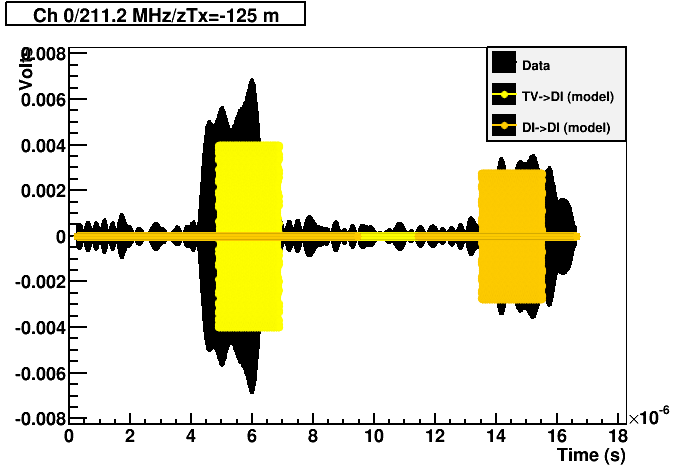}
\includegraphics[width=0.48\textwidth]{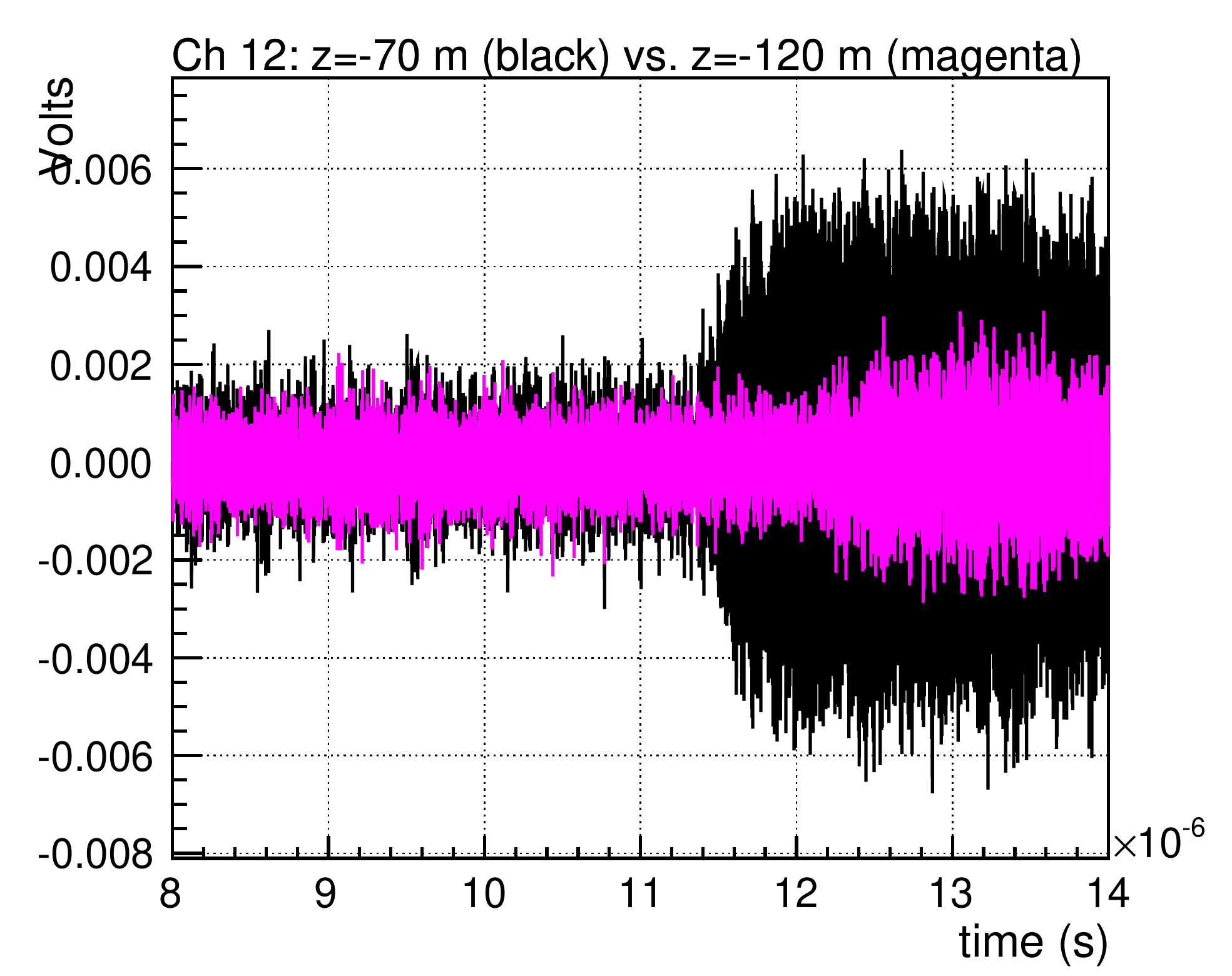}
\caption{Left: Overlay of raw data obtained in 2003 horizontal propagation experiments with model predictions for $TV\to DI$ (light yellow) and also $DI \to DI$ (dark yellow).
Right: Comparison of $DI \to DI$ signal arrival times in RICE channel 12 ($z=-110$ m) for transmitter in NOAA borehole, at depths $z=-70$ m (black) vs.~$z=-120$ m (magenta; scaled), illustrating time delay of latter relative to former.}
\label{fig:XSPmodel}
\end{figure}

To determine the possible ray trajectories consistent with the observed timing of these signals, we compare these data with simplified models of ray propagation. Owing to uncertainties in the surface elevation map, and thus the point-of-entry into the ice for the $TV \to DI$ path, our model for the $TV \to DI$ ray is `extreme' (and unphysical) - namely, we calculate the expected timing for a ray traveling horizontally through air, then bending ninety degrees into the ice to the in-ice receiver. Nevertheless, this model differs from the expected signal trajectory by only $\cal{O}$(100 ns) in transit time, which is commensurate with the magnitude of our overall total timing uncertainties. For the $DI \to DI$ path, we use the ARA Collaboration experimental model \citep{abdul2017measurement} for the index-of-refraction profile, which we integrate, assuming straight-line ray propagation from source in-ice dipole transmitter to in-ice receiver, to determine the total transit time.
These predictions are overlaid with data in Fig.~\ref{fig:XSPmodel}. In general, our very crude model matches data to within $\sim$0.5 $\mu$s, commensurate with the sum of uncertainties due to ray trajectory, cable delays, trigger time delays, etc. Expressed fractionally, that uncertainty
is $\sim$5\% on the total travel time, or $\sim$0.02--0.03 in the refractive index.

Fig.~\ref{fig:XSPmodel} (right) overlays the signal arrival for the case where the transmitter is at a depth of $-70$ m vs.~$-120$ m. We observe $\sim$0.8 $\mu$s time delay stagger in the latter relative to the former, compared with $\sim$0.75 $\mu$s assuming least-time propagation to the RICE channel 12 receiver ($z=-110$ m). This observed time delay difference is incompatible with through-air or surface signal propagation from the transmitter at the two depths (z=--70 m and z=--120 m) to the receiver, which would imply a much shorter time stagger in their received signals of no more than 0.2 $\mu$s. We also note an extended period of signal onset, indicating a wide range of contributing ray trajectories, consistent with the observation that many of the received $DI\to DI$ signals are apparently temporally broadened compared to the nominal tone signal duration.

\subsubsection{Estimate of attenuation length from shadowed transmissions}
\begin{figure}\centerline{\includegraphics[width=0.6\textwidth]{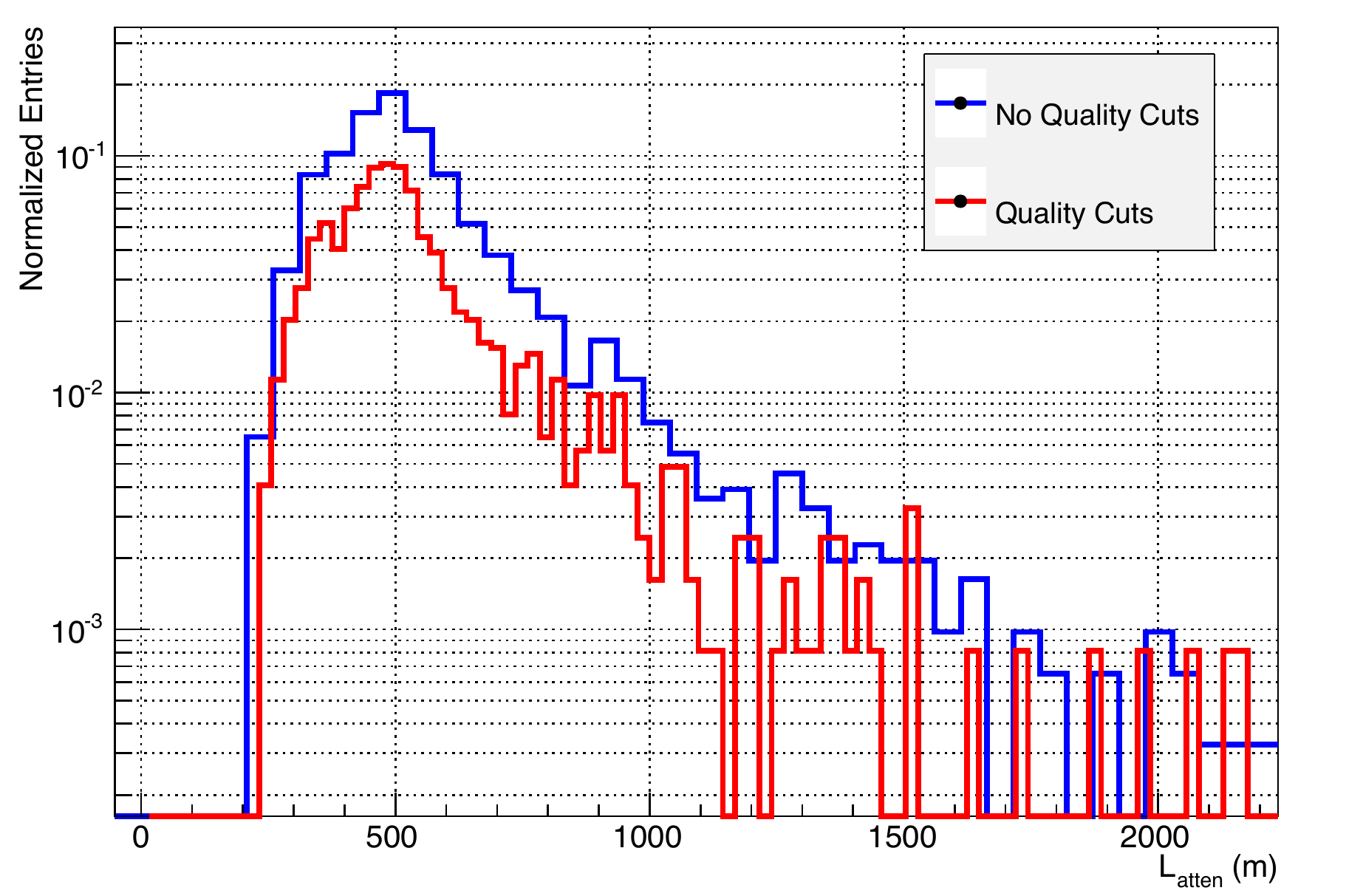}}
\caption{Distribution of field attenuation lengths calculated using RICE signal amplitudes measured from `far' transmitter relative to `near' transmitter. ``Quality Cuts'' refers to restricting data sample for which receiver Signal-to-Noise Ratio (SNR) exceeds 6:1 in amplitude.}
\label{fig:XSPLatten}
\end{figure}

Given multi-channel RICE receiver data, and using multiple data runs taken at both the near and far locations to sample a variety of depths and frequencies as well as a range of possible systematics, the attenuation length for horizontal propagation can be calculated by normalizing the signal strengths measured, channel-by-channel and run-by-run, for broadcasts over 3.3 -- 3.5 km baselines, to signal strengths measured, channel-by-channel and run-by-run, to broadcasts when the transmitter is located within the RICE array itself (``near'' transmission). Assuming the simplest 1/r electric field dependence, we apply corrections for the distance difference between the near and far locations, and also for the $\cos\theta$ dependence of the dipole beam pattern, as outlined previously. 
The ensemble of electric field attenuation lengths extracted in this manner is presented in Fig.~\ref{fig:XSPLatten}. As a systematic check, we have sub-divided our samples by depth of the transmitter at the far location and also frequency of signal broadcast (Table \ref{tab:z}). Our observed scatter in calculated attenuation lengths is consistent with our estimated systematic errors. In principle, possible dispersive effects for shadow propagation can be probed by measuring the signal onset time, relative to the $TV\to TV$ trigger, over the frequency range probed in this experiment. In practice, uncertainties in such a measurement were comparable to the determination of the signal onset time, and must therefore await more precise future measurements.

\begin{table}
\begin{center}
\caption{Observed signal-to-noise ratio and calculated average field attenuation length dependence on transmitter depth (top, and summing over all frequency data), and on frequency (bottom, and summing over all transmitter depth values); statistical errors only are shown. Estimated systematic errors are comparable in magnitude to the spread observed in the data points.}
\begin{tabular}{|c|cc|}
\hline
$z_{Tx}$ & $\langle SNR \rangle$ & $L_{atten}$ (m) \\ 
\hline
70  &    13.9$\pm$1.9 & $521\pm12.2$ \\
120 &   11.4$\pm$0.8 & $476\pm8.5$  \\
125 &   13.4$\pm$1.3 &   $491.2\pm9.8$ \\
\hline
\end{tabular} \\
\vspace{0.5cm}
\begin{tabular}{|c|c|}
\hline
 frequency (MHz) & $L_{atten}$ (m) \\ 
\hline
 211.2 & 484.4$\pm$4.5 \\
 230 & 495.7$\pm$2.8 \\
 490 & 563.2$\pm$27.3\\
\hline
\end{tabular}
\label{tab:z}
\end{center}
\end{table}

The ice is expected to have complete horizontal translational symmetry, with vertical symmetry broken by the presence of conducting layers within the ice (primarily due to deposits following volcanic eruptions) and/or fluctuations in the vertical density profile. As noted earlier, such vertical asymmetries suggest models in which signal emitted isotropically might be `trapped' in a horizontal channel, thus circumventing the otherwise-expected shadowing. We have therefore searched for a possible inverse dependence of the calculated attenuation length on the vertical separation between transmitter and receiver. Our data suggest a possible slight decrease in attenuation length with the magnitude of $z_{Tx}-z_{Rx}$, although insufficient to be conclusive. 

Numerically, our extracted attenuation length for all possible near/far combinations (550$\pm$10 m, where the error shown is the error on the mean) is consistent with the result obtained when we restrict our calculation to those `high-quality' combinations having high signal-to-noise only (542$\pm$16 m).


We also note that similar broadcasts from the 8-km distant SPRESO hole (South Pole Remote Earth Science and Seismological Observatory), with transmitter at z=-300 m,  yielded no observable signal in the RICE channels. This is
consistent with the large number
of implied e-foldings ($\sim$16) to the RICE receiver array by a 550 m attenuation length. Unshadowed propagation from that source point should have yielded SNR values approximately 2--3$\times$ larger than those observed from the NOAA source location.

\subsubsection{Cross-checks and Possible Systematic Errors}
For the RICE measurements described herein, the uncertainty in the signal arrival times is estimated as approximately one time sampling bin (i.e., one nanosecond), which is insignificant compared to the $\sim$10 microsecond total travel times. Uncertainties in the attenuation length measurement are reflected in the width of the distribution shown in Fig.~\ref{fig:XSPLatten}, or approximately 25 meters. 
Additional cross-checks were made to ensure that signal was
not otherwise being lost in the signal path from generator to in-ice dipole, including checks for: a) faulty cables, connectors, antennas, or amplifiers (checked by swapping in/out other cables, connectors, antennas, or amplifiers), b) non-linearity of the power amplifier to the in-field transmitter, which was checked by direct measurement, c) sensitivity to possible coupling of the in-ice dipole antennas to the sides of the borehole, which was checked by taking multiple measurements after successively entirely raising and lowering the transmitter dipole, and d) saturation of the near-hole receiver amplifiers, which was also checked by verifying the linearity of the received near-hole signals with transmitter gain.

\subsection{Observation of horizontal propagation with the ARIANNA experiment at Moore's Bay, Ross Ice-Shelf, Antarctica}
After earlier prototypes, deployment of the pilot-stage ARIANNA Hexagonal Array (HRA) began in 2014 and has since demonstrated successful operation under harsh Antarctic conditions \citep{Barwick:2014rca}. ARIANNA employs high-gain log-periodic dipole antennas (\emph{LPDAs}) with excellent broad-band response between 100 MHz and 900 MHz, primarily sensitive to signals polarized parallel to the antenna tines. ARIANNA comprises multiple \emph{stations}, each acting as an independent autonomous neutrino and cosmic ray detector and including four (or more) LPDAs deployed just below the snow surface, admitting easy access and repair when necessary. 
The sensitivity to radio signals from neutrino interactions is enhanced by the high dielectric contrast at the ice-water interface at the bottom of the Ross Ice-Shelf, resulting in efficient reflection of down-going emission back towards the antennas \citep{neal_1979,hansJGlac}.

The first installed HRA stations have been used to derive limits on the neutrino flux \citep{Barwick:2014pca} and to measure the radio emission of air showers, which are an important background for arrays with antennas close to the surface, while simultaneously providing equally important proof-of-concept and calibration \citep{barwick2017radio}. ARIANNA measurement of air showers has demonstrated that the hardware response of the experiment, including antennas and amplifiers, is well-understood, as the predicted signal shape is well-matched by the detected signal shapes \citep{barwick2017radio}. ARIANNA is, thus far, the only ground-based experiment to successfully self-trigger on radio emissions from air showers, with a high purity and efficiency independent of particle detectors. 

The deployment of the HRA was accompanied by several ice properties measurements in subsequent years \citep{hansJGlac}. For many tests, such as studies of signals reflected off the ice-water interface on the bottom of the ice-shelf, early signals were measured in stations for which in-ice propagation was believed to be forbidden by the shadowing effect. These were initially not emphasized because they were considered as either potential in-air propagation or artifacts of the measurement set-up \citep{JordanHansonUCIphdthesis}. Additional analysis ruled out those possibilities, indicating that such signals were likely due to horizontal propagation, putative shadowing notwithstanding, prompting recent direct measurements of this phenomenon.

\subsubsection{Experimental Technique}
\begin{figure}
\centerline{\includegraphics[width=0.6\textwidth]{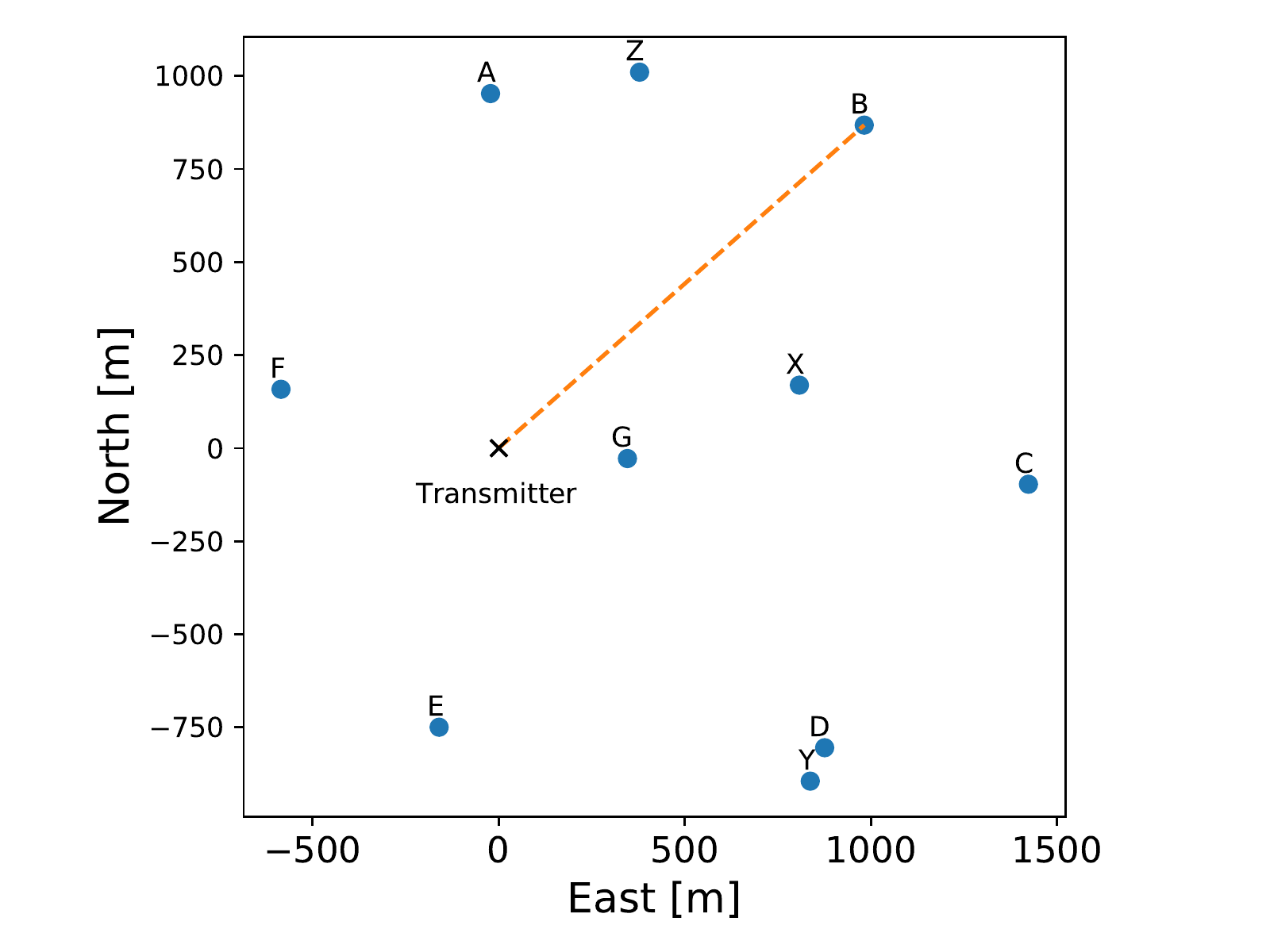}}
\caption{ARIANNA HRA array in the Antarctic season 2016/17. The transmitter for these studies was positioned at the origin in these coordinates. An example path of propagation to the station at position B is indicated by the dashed line.}
\label{ARIANA:array}
\end{figure}

During the 2016-17 polar season, two dedicated boreholes, separated by about 100 meters, were drilled to a depth of 20 meters in the ice, to permit the detailed study of horizontally propagating signals. (In practice, snow infall in the hole resulted in data taken at $z=-19$ m, rather than $z=-20$ m.) While transmitting and receiving between the two boreholes, multiple ARIANNA stations, schematically outlined in Fig.~\ref{ARIANA:array}, were also regularly recording data, allowing for redundant cross-checks of propagation over multiple baselines. Signals were broadcast from the same RICE fat-dipoles used in the 2003 RICE study described previously.

\subsubsection{Measurements between boreholes}
High-amplitude (few kV), short-duration signals ($<20$ ns) were generated using a Pockels Cell Driver (\emph{PCD}), which was routed directly to the transmitting antenna located in one borehole.  Absolute timing was achieved by using a BNC Model 555 pulser to simultaneously trigger the PCD and send a triggering pulse to an oscilloscope for monitoring the signals received in the other borehole.

For reference, the entire set-up was lifted into the air, both on the ice-shelf and also pre-deployment in a park in California (i.e. dry ground, very little conductivity, flat area, little high vegetation); recorded signals were observed to be of similar strength at both locales. 

\begin{figure}
\includegraphics[width=0.4\textwidth]{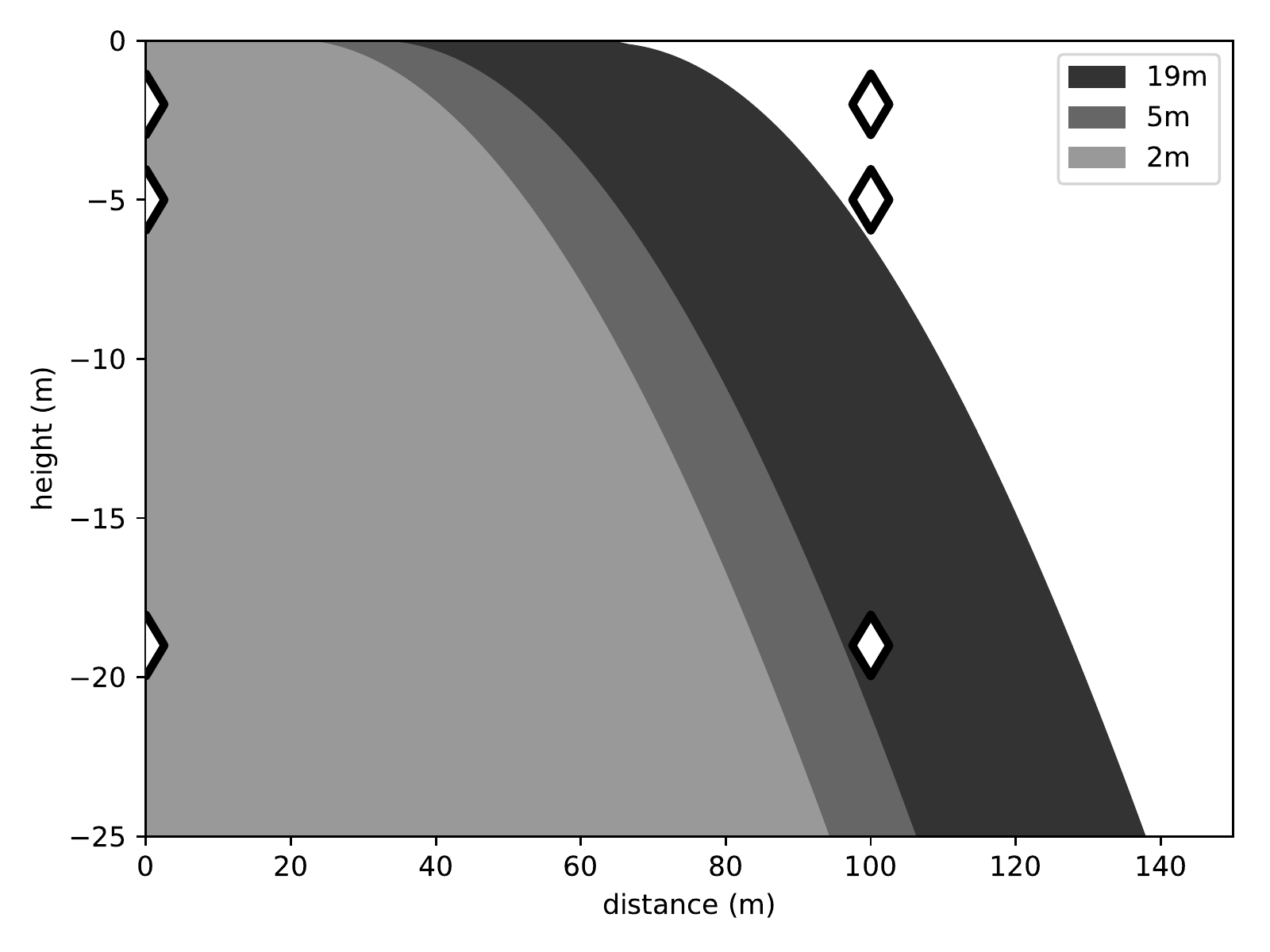}
\includegraphics[width=0.55\textwidth]{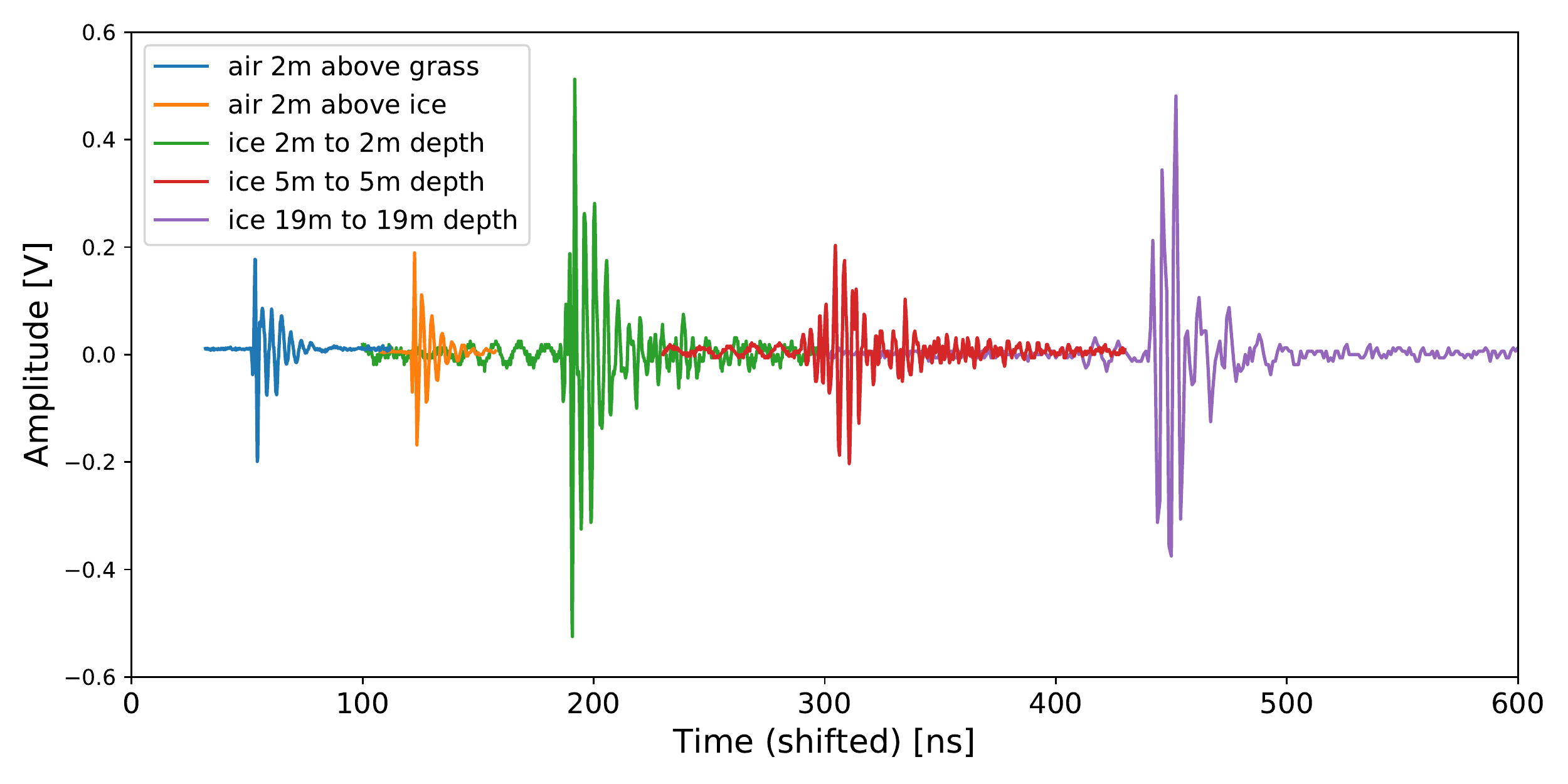}
\caption{Left: Illustration of shadowing at the ARIANNA site. Shaded regions indicate the horizon visible to a receiver (right) at the indicated depths of each transmitter (left). Diamonds show the location of the different transmitters and receiver positions in the first and second bore-hole, respectively. Right: Pulses as recorded when transmitting from a RICE dipole to a receiving RICE dipole at varying transmitter/receiver depths at Moore's Bay. For reference measurements in air (in Moore's Bay and in California) are also depicted. Pulses have been arbitrarily shifted horizontally to better illustrate signal shapes.}
\label{illustration}
\end{figure}

For all tested in-ice configurations of different depths, strong pulses are observed.
This is despite the fact that simple ray tracing would only allow for signals in certain combinations as shown on the left in Fig.~\ref{illustration}. The Figure
also shows sample pulses recorded in different configurations of Tx/Rx. The signal shapes are observed to be very similar, despite the fact that, absent shadowed propagation, only the in-air configurations and the Tx/Rx depth of 19 meters should be visible. Interestingly, the amplitudes vary quite significantly, despite no changes in the set-up. This can at this point only tentatively be attributed to multi-path effects which lead to constructive and destructive interference.

We note that a) signal timing is consistent with horizontal propagation (as we quantify below), and b) no special transition was experimentally observed when the transmitter was moved across the shadow/non-shadow zone boundary.

\begin{figure}
\centering
\includegraphics[width=0.55\textwidth]{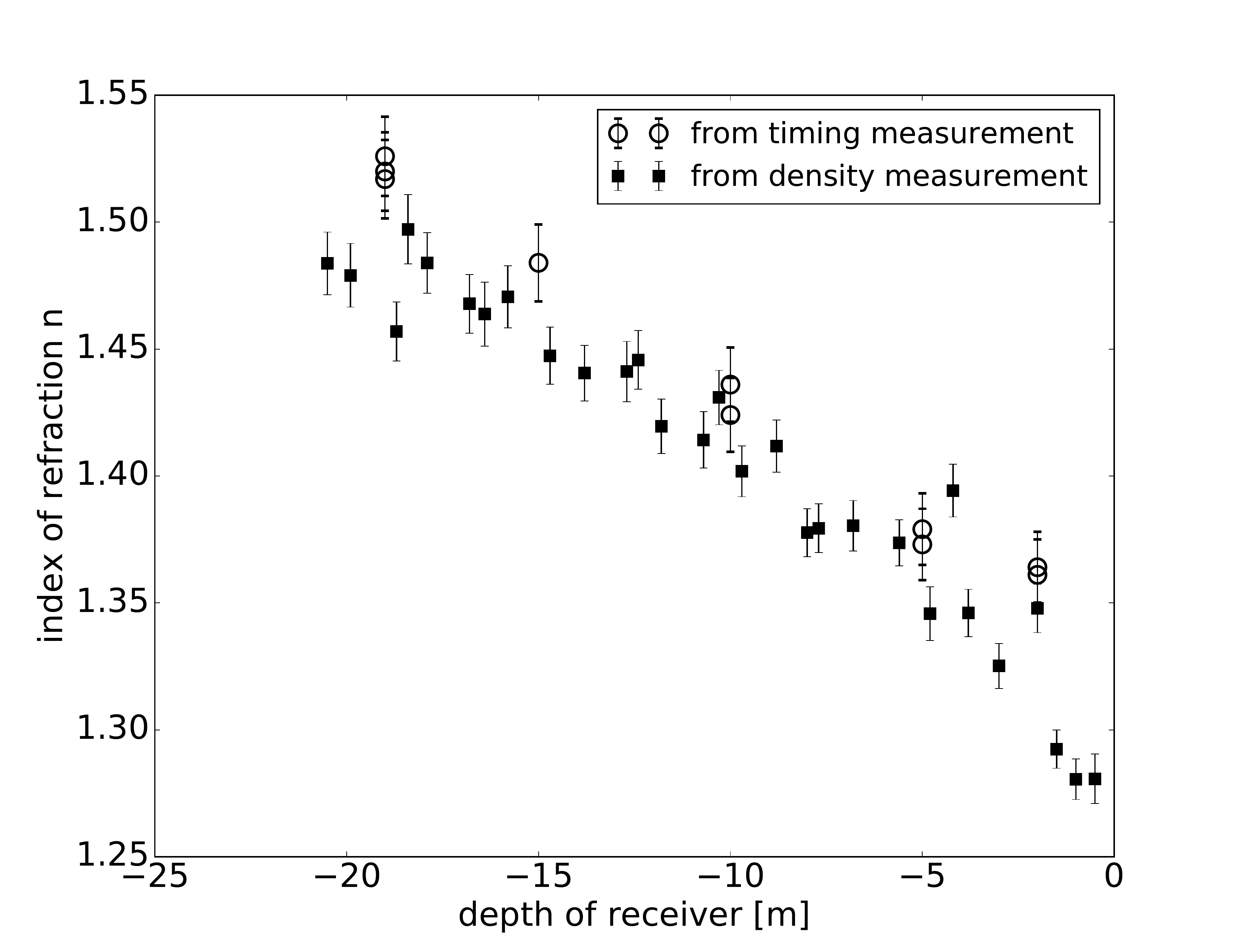}
\caption{Measured index of refraction as function of depth. Filled squares indicate the index of refraction as calculated from ice density measurements of the bore holes. The open circles show the index of refraction calculated from absolute timing of the propagation of the pulses, using the leading edge of each pulse to obtain signal arrival times. 
}
\label{indexofrefr_1}
\end{figure}

In order to confirm that the observed pulses propagate through ice vs.~air, the average index of refraction was measured for every combination of dipole depths from the signal arrival times, and then compared to the index-of-refraction calculated from ice density measurements obtained during hole-drilling. As Fig.~\ref{indexofrefr_1} shows, the timing is fully compatible with propagation through the ice, and incompatible with through-air propagation ($n\approx 1.0$), for a variety of depths. It is also incompatible with the ice-water boundary bounce hypothesis, as the measured timing cannot be reconciled with two-way propagation through the ice-shelf ($\sim 1000$ m) and an index of refraction $n>1.0$.

It should be noted that there seems to be a systematic offset between the index of refraction derived from the timing measurements and the ice density measurements (perhaps resulting from multi-path effects). As the index of refraction obtained from the air$\to$air measurement (n=1.016) is approximately 1.6\% higher than expectation, this offset may also be a systematic effect. To exclude the possibility that signals were the result of accidental emission of the PCD itself, it was also verified that no signals were observed when the transmitting antenna in the ice was disconnected from the PCD. 

The pulses received in the shadow zone (Tx 19 to Rx 2)  have an average signal-to-noise ratio (SNR) of 25. Albeit being longer than the pulses received in the allowed zone (SNR = 193) the signals are still well-above the noise floor and contain significant power.

\subsubsection{Measurements in ARIANNA stations}
\begin{figure}
\centerline{\includegraphics[width=0.7\textwidth]{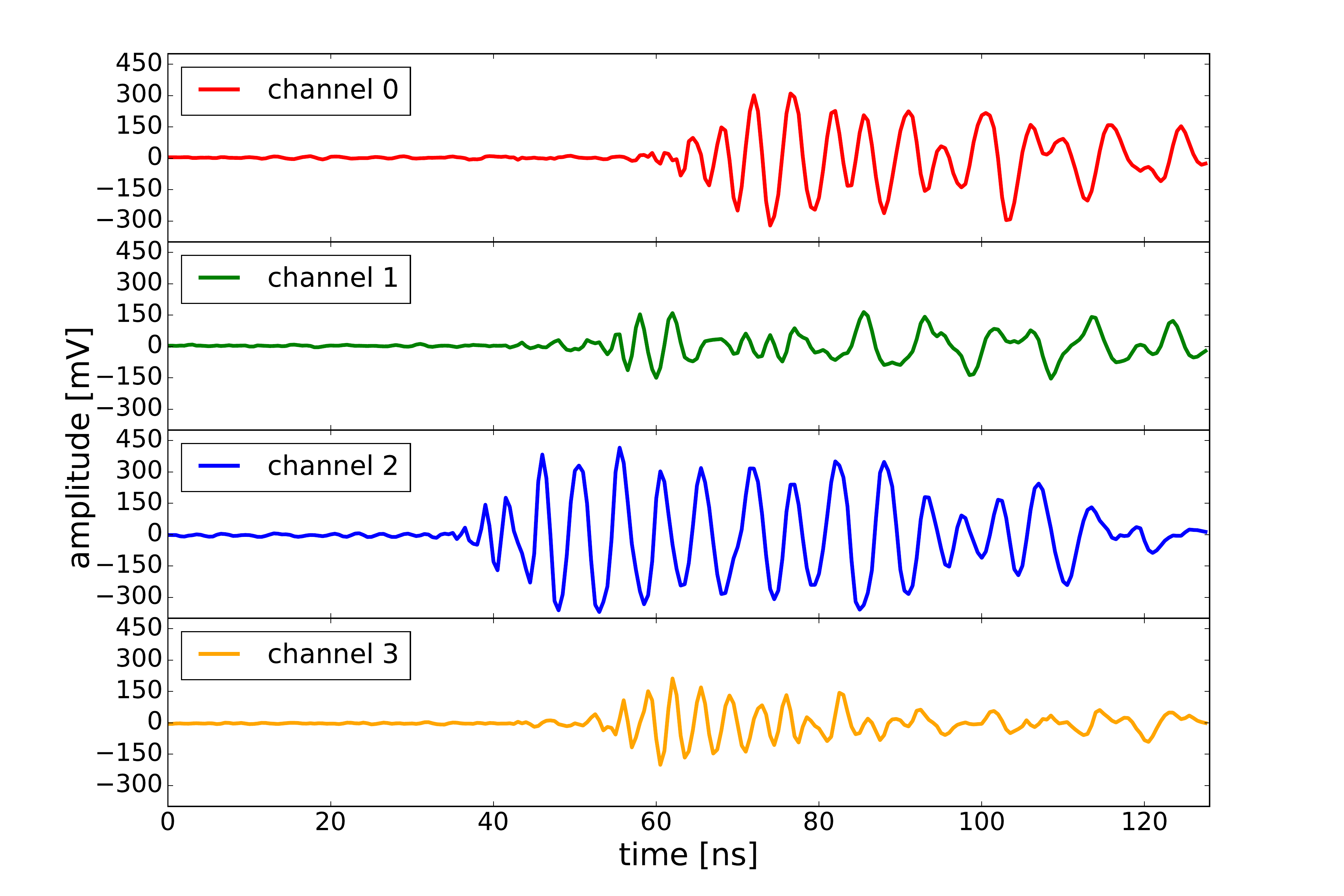}}
\caption{Horizontally propagating signals recorded in ARIANNA station \emph{A}, at a horizontal distance of 953 meters from the transmitter. All pulses show a sharp leading edge with signal persisting for tens of nanoseconds, possibly extending beyond the length of the ARIANNA waveform record. The antennas of channels 0 and 2 are aligned roughly perpendicular to the arrival direction from the transmitter, while channels 1 and 3 are almost parallel. Channel 2 is closest to the transmitter and channel 0 furthest away, with a difference in distance of about 6 meters.}
\label{fig:station19}
\end{figure}

While pulsing in the boreholes, signals were captured in all normally operational ARIANNA stations with their nominal trigger settings. Fig.~\ref{fig:station19} shows signals as recorded in a station at a distance of 953 meters from the transmitter, for which shadowing would otherwise prohibit signal observation. Since there is no absolute timing information between the transmitter and the ARIANNA station, to demonstrate that observed radio signals are propagating horizontally and are not the result of reflections from the underlying Ross Sea-Ross Ice Shelf boundary, one can consider arrival times within a station. Every HRA station is equipped with two co-polarized pairs of 6-meter separated LPDAs, with different pairs oriented perpendicularly, allowing direct polar angle-of-incidence inference based on a single antenna pair. 

\begin{figure}
\centering
\includegraphics[width=0.5\textwidth]{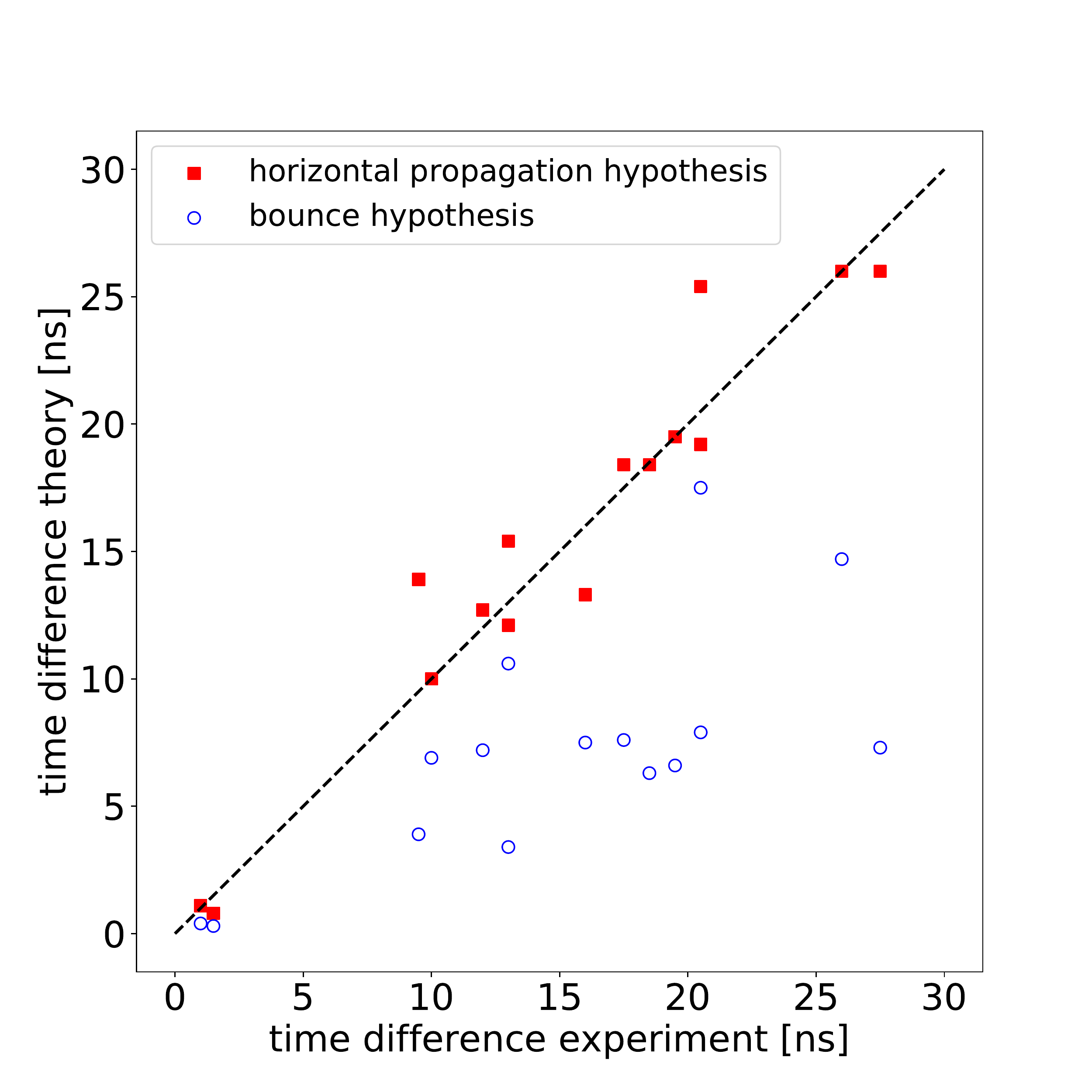}
\caption{Measured (x-axis) vs.~expected (y-axis) signal arrival times in ARIANNA stations for two different propagation hypotheses. The filled squares correspond to the \emph{bounce}-hypothesis, while the open circles represent the hypothesis of horizontal propagation. The proximity to the dashed line through the origin with slope one illustrates that the arrival times match horizontal propagation through the ice better than a reflection from the underlying Ross Sea. }
\label{indexofrefr_2}
\end{figure}

Fig.~\ref{indexofrefr_2} shows the contrast between expected and measured arrival directions when using the \emph{bounce} hypothesis vs.~the horizontal propagation hypothesis. For both hypotheses the time difference between pulses in antenna pairs are calculated (y-axis) and compared to the measured time difference (x-axis). A good agreement is reached, when the points follow the dashed line through the origin and prediction matches measurement.  

While there is significant scatter (partly due to the rather simple method chosen to identify the timing of the signal, as well as the short waveform length), the observed signals clearly favor horizontal propagation. We also observe no strong polarization dependence in those received signals.

The time structure of the pulses suggests some dispersion, as the received signal is elongated in time as compared to the emitted signal (Fig.~\ref{fig:station19}). 
Some dispersion is expected from the antenna and amplifier response of the ARIANNA stations \citep{Barwick:2014boa}. Amplifiers with a small group-delay are difficult to accommodate in low-power, broadband systems and the LPDAs are also slightly dispersive due to their broadband nature. However, even assuming the least sensitive direction for the LPDAs (a fully vertically polarized signal arriving in the null of the antenna i.e. parallel to the tines) cannot account for the dispersion of the signal observed. 

Dispersive effects have the negative consequence of stretching the signal in the time domain and thereby reducing the instantaneous amplitude, but also offer the possibility that the relative arrival time of different frequency components can provide information on the distance-to-vertex for future in-ice neutrino searches, which is essential for a neutrino energy estimate. 

\begin{figure}
\centerline{\includegraphics[width=0.68\textwidth]{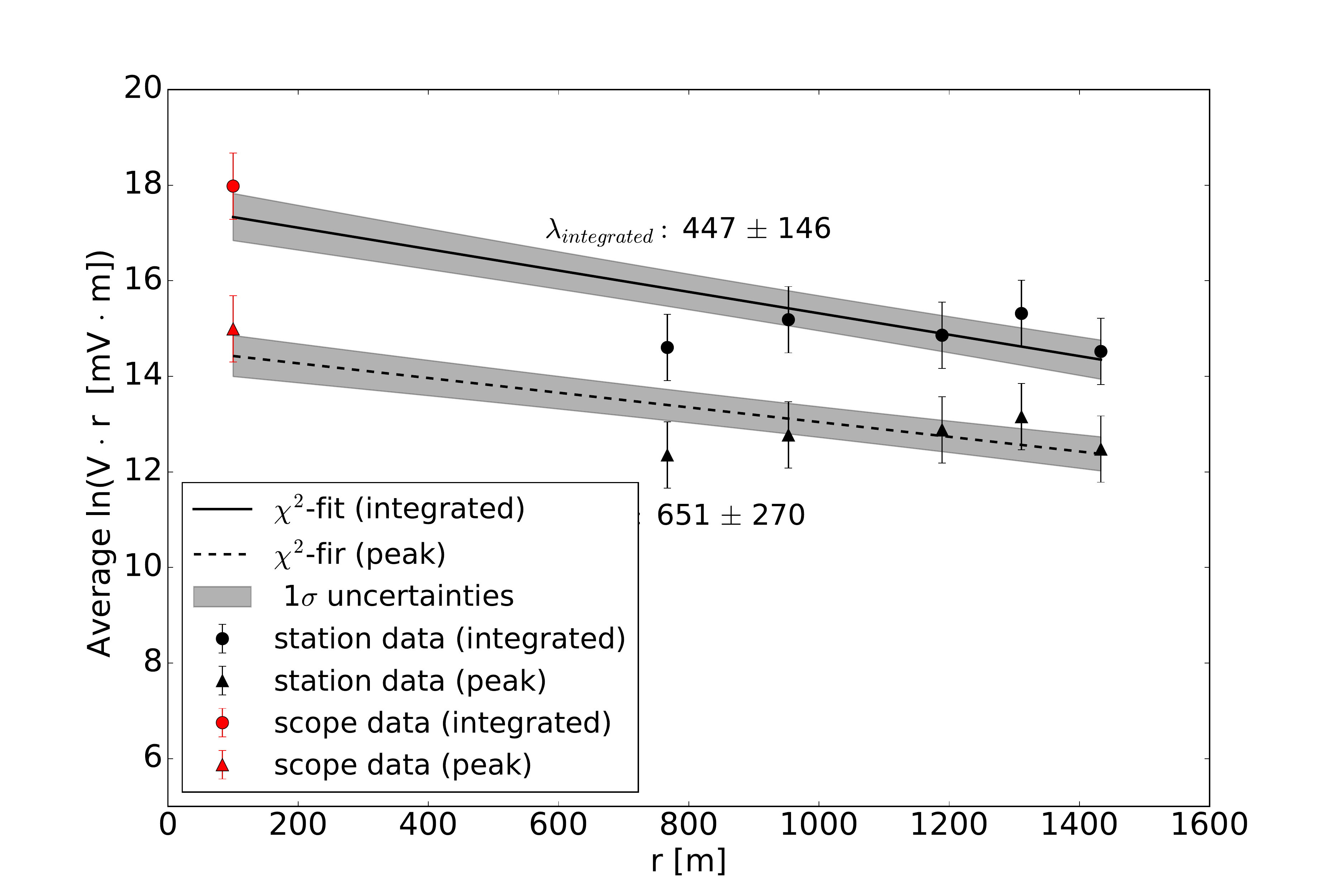}}
\caption{Field attenuation derived from all measured signals. Two calculations are made, one using the peak amplitudes of the signals and one using integrated absolute amplitudes (150 - 250 MHz), to account for possible dispersion. The reconstructed attenuation lengths from the pulse amplitude is $651 \pm 270$ meters; for the integrated amplitude, the corresponding value is $447\pm146$ meters. Note that exclusion of the left-most data point, which has been measured with an oscilloscope and not a station and therefore might be subject to different uncertainties, results in an even longer estimated attenuation length.}
\label{attenuation}
\end{figure}

The electric field attenuation lengths extracted from the data collected with all ARIANNA stations and the neighboring borehole are compiled in Fig.~\ref{attenuation}. Two calculations have been made to cross-check whether there is a significant difference when accounting for possible dispersive effects. 
Neglecting differences in the systematic uncertainties between the data obtained with an oscilloscope and ARIANNA station data, the best fit results in an attenuation length of $447\pm 146$ meters for the time-integrated absolute amplitude and $651\pm 270$ meters based on the peak observed pulse amplitude only. These values are commensurate with those obtained at South Pole for horizontal propagation. These results imply that 1/r geometric signal reduction dominates over attenuation for horizontal propagation, which is an intriguing perspective for the effective volume for an ARIANNA-like detector with a station spacing of about one kilometer. 

\subsubsection{Measurement from single borehole to buried LPDA}
The same set-up with the PCD was also used to recorded pulses between the transmitter in a borehole and an LPDA buried at a depth of one meter in December of 2017. The LPDA was placed at a distance of 500 meters, which corresponds to the longest signal cable available. The tines of the LPDA were rotated perpendicular to the line connecting its position and the borehole for maximum gain. Using a cabled set-up and the long record of an oscilloscope allows for absolute timing. 

\begin{figure}
\centering
\includegraphics[width=0.8\textwidth]{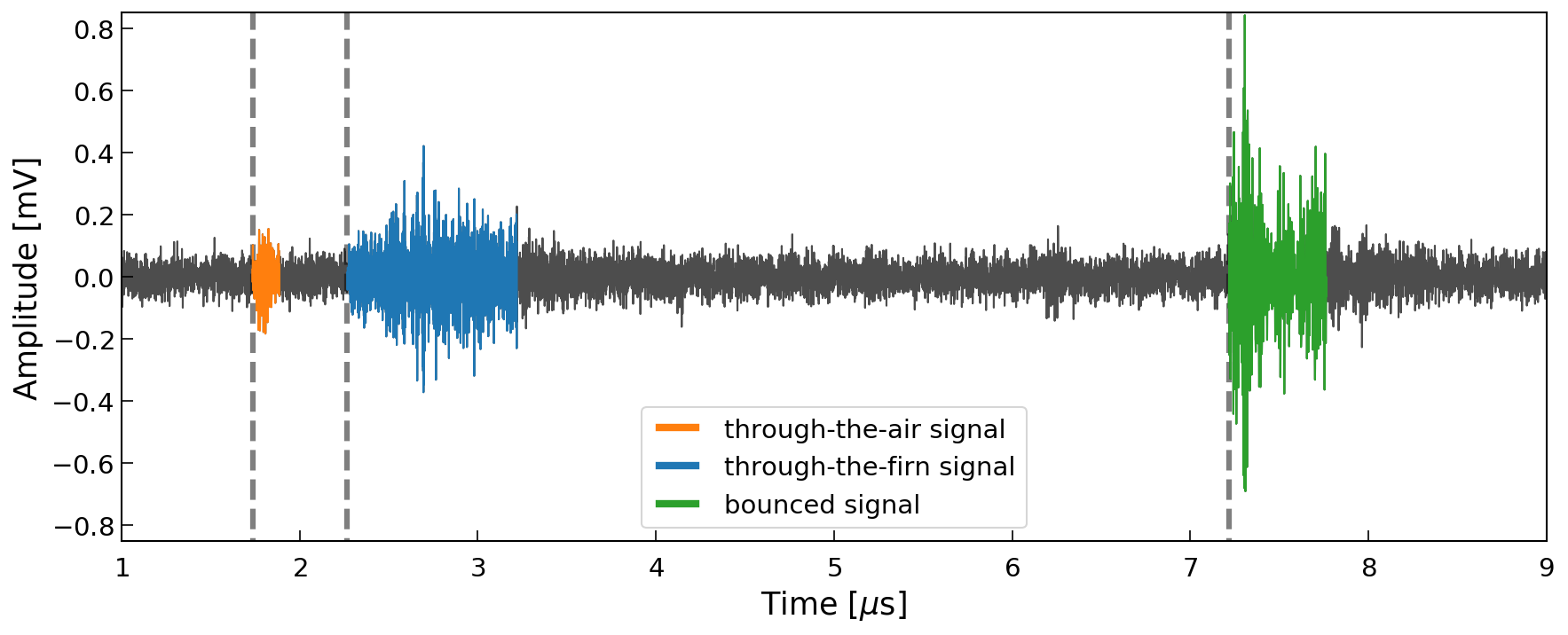}
\caption{Pulses recorded with an LPDA buried at one meter depth, transmitted from a dipole at 19 meters depth at 500 meters distance. The dashed lines correspond to the calculated travel times in air ($n=1.0$, + propagation up through firn), firn (n=$1.36$) and the ray tracing solution for a pulse reflecting from the bottom of the ice-shelf, assuming a thickness of  578 meters.}
\label{pulseswithtiming}
\end{figure}

Three pulses were observed as shown in Fig.~\ref{pulseswithtiming}. Solely from timing, the three pulses can be attributed to different paths between transmitter and receiver. The first small pulse has to travel (mainly) through the air, as its arrival time corresponds to a propagation with the speed of light in air ($n = 1.0$). The signal is no longer present when the Tx antenna is disconnected from the PCD, so the signal is emitted by the antenna and has to propagate up and out of the firn first and then along the surface. The second pulse is compatible with the horizontal propagation through the firn, as its start-time corresponds to a propagation through a medium having $n=1.36$. At a distance of 500 meters this horizontal propagation is not an allowed solution of classical ray tracing. The third pulse is found at the time required to travel twice through the ice-shelf and is therefore the reflection of the original signal off the bottom of the ice-shelf. 

The Figure depicts the raw data waveforms recorded in the field, with no applied gain correction. As an LPDA is rather insensitive to signals arriving perpendicular to the plane formed by the dipole elements (at least 3 dB compared to its front-lobe), the signals arriving horizontally are suppressed in this measurement, and may well contain more power than the reflected signal. An exact quantification requires knowledge of the precise arrival direction and the polarization of the incoming signal, which is impossible with the single LPDA which has been used to conduct these measurements. 
The reported signal strength is therefore a lower limit on the true power in the horizontally propagating signal.   

Additional data, taken during the 2017-18 Antarctic field season, is currently being analyzed and should improve the understanding of signal propagation at Moore's Bay. Additional studies, focusing of signal polarization, are foreseen for the 2018-19 season. 

\section{Conclusions}
We have presented evidence for electromagnetic signals propagating from nominally shadowed source locations. 
Although forbidden in the ray optics picture for the case of a smooth and monotonic variation of wave-speed with depth, reflective layers or local deviations from the smooth n(z) profile can result in local signal channeling. 

The measured attenuation length of $\sim$500 m, from both the South Polar and the Ross Ice Shelf locales, is slightly larger than the average unshadowed attenuation length measured at Moore's Bay \citep{hansJGlac}, and approximately one-third that observed for unshadowed radio signal propagation at South Pole \citep{allison2012design} in the upper 1.5 km of ice. This value is also compatible with what might be expected in a model where microscopic scattering occurs entirely incoherently, and phase information is lost in the scattering process. 

The attenuation length for horizontal propagation presented herein is comparable to the maximum detectable range for $\sim$10 PeV neutrinos using the radio technique. Contrary to previous expectation, experimental observation of such neutrinos is therefore not limited by shadowing. This neutrino energy regime is particularly interesting experimentally, as it represents the maximum upper energy reach of the IceCube experiment and the detected astrophysical neutrino flux. A radio detector with such an energy threshold may measure the continuation of the IceCube flux, which is likely orders of magnitude larger than the flux caused by the interaction of cosmic rays with the cosmic microwave background at 100 PeV.
As the horizontally propagating signals are well above the noise floor, this result therefore suggests that a future neutrino detector constructed at relatively shallow depths ($<30$ meters) might optimize the balance between science return and the logistical overhead associated with hole drilling.

During the 2017-18 austral season, a transmitter based on the HiCal \citep{gorham2017hical} piezo-electric model was lowered into the SPICE core hole \citep{casey20141500} and broadcast to both the ARA array (deep) as well as a single ARIANNA station at the surface, allowing a more systematic map of signal transmission over a range of depths. Data analysis is currently in progress. A second generation of those studies will be conducted in 2018-19. 

\section{Acknowledgments}
 We are grateful to the U.S. National Science Foundation-Office of Polar Programs, the U.S. National Science Foundation-Physics Division and the U.S. Department of Energy.  
We thank generous support from the German Research Foundation (DFG), grant NE 2031/1-1, NE 2031/2-1 and GL 914/1-1, the Taiwan Ministry of Science and Technology, the ``Wallenbergstiftelsen'' (Wallenberg Foundation) and the ``Liljewalch stipendier'' (Liljewalch scholarships).

\section*{References}
\bibliographystyle{model1-num-names}
\bibliography{Zref}

\begin{thebibliography}{30}
\expandafter\ifx\csname natexlab\endcsname\relax\def\natexlab#1{#1}\fi
\providecommand{\bibinfo}[2]{#2}
\ifx\xfnm\relax \def\xfnm[#1]{\unskip,\space#1}\fi
\bibitem[{{Aartsen} et~al.(2016)}]{2016ApJ...833....3A}
\bibinfo{author}{M.~G. {Aartsen}}, et~al.,
\newblock \bibinfo{title}{{Observation and Characterization of a Cosmic Muon
  Neutrino Flux from the Northern Hemisphere Using Six Years of IceCube Data}},
\newblock \bibinfo{journal}{The Astrophysical Journal} \bibinfo{volume}{833}
  (\bibinfo{year}{2016}) \bibinfo{pages}{3}.
\bibitem[{Askaryan(1962)}]{Askaryan1962a}
\bibinfo{author}{G.~A. Askaryan},
\newblock \bibinfo{title}{{Excess negative charge of an electron-photon shower
  and its coherent radio emission}},
\newblock \bibinfo{journal}{Soviet Physics JETP} \bibinfo{volume}{14}
  (\bibinfo{year}{1962}) \bibinfo{pages}{441--443}.
\bibitem[{{Askaryan}(1962)}]{Askaryan1962b}
\bibinfo{author}{G.~A. {Askaryan}},
\newblock \bibinfo{title}{{Excess negative charge of electron-photon shower and
  the coherent radiation originating from it. Radiorecording of showers under
  the ground and on the moon}},
\newblock \bibinfo{journal}{J. Phys. Soc. Japan} \bibinfo{volume}{Vol. 17,
  Suppl. A-III} (\bibinfo{year}{1962}) \bibinfo{pages}{257}.
\bibitem[{{Askaryan}(1965)}]{Askaryan1965}
\bibinfo{author}{G.~A. {Askaryan}},
\newblock \bibinfo{title}{{Coherent Radio Emission from Cosmic Showers in Air
  and in Dense Media}},
\newblock \bibinfo{journal}{Soviet Phys. JETP} \bibinfo{volume}{21}
  (\bibinfo{year}{1965}) \bibinfo{pages}{658}.
\bibitem[{Barrella et~al.(2011)Barrella, Barwick, and
  Saltzberg}]{barrella2011ross}
\bibinfo{author}{T.~Barrella}, \bibinfo{author}{S.~W. Barwick},
  \bibinfo{author}{D.~Saltzberg},
\newblock \bibinfo{title}{{Ross Ice Shelf (Antarctica) in situ radio-frequency
  attenuation}},
\newblock \bibinfo{journal}{Journal of Glaciology} \bibinfo{volume}{57}
  (\bibinfo{year}{2011}) \bibinfo{pages}{61--66}.
\bibitem[{Barwick et~al.(2005)Barwick, Besson, Gorham, and
  Saltzberg}]{barwick2005south}
\bibinfo{author}{S.~W. Barwick}, \bibinfo{author}{D.~Besson},
  \bibinfo{author}{P.~Gorham}, \bibinfo{author}{D.~Saltzberg},
\newblock \bibinfo{title}{{South Polar in situ radio-frequency ice
  attenuation}},
\newblock \bibinfo{journal}{Journal of Glaciology} \bibinfo{volume}{51}
  (\bibinfo{year}{2005}) \bibinfo{pages}{231--238}.
\bibitem[{{Gorham} et~al.(2009)}]{GorhamAllisonBarwick2009}
\bibinfo{author}{P.~W. {Gorham}}, et~al.,
\newblock \bibinfo{title}{{The Antarctic Impulsive Transient Antenna ultra-high
  energy neutrino detector: Design, performance, and sensitivity for the
  2006-2007 balloon flight}},
\newblock \bibinfo{journal}{Astropart. Phys.} \bibinfo{volume}{32}
  (\bibinfo{year}{2009}) \bibinfo{pages}{10 -- 41}.
\bibitem[{Barwick et~al.(2015)}]{Barwick:2014pca}
\bibinfo{author}{S.~W. Barwick}, et~al.,
\newblock \bibinfo{title}{{A First Search for Cosmogenic Neutrinos with the
  ARIANNA Hexagonal Radio Array}},
\newblock \bibinfo{journal}{Astropart. Phys.} \bibinfo{volume}{70}
  (\bibinfo{year}{2015}) \bibinfo{pages}{12--26}.
\bibitem[{{Kravchenko} et~al.(2003){Kravchenko}, {Frichter}, {Seckel},
  {Spiczak}, {Adams}, {Seunarine}, {Allen}, {Bean}, {Besson}, {Box}, {Buniy},
  {Drees}, {McKay}, {Meyers}, {Perry}, {Ralston}, {Razzaque}, and
  {Schmitz}}]{KravchenkoFrichterSeckel2003}
\bibinfo{author}{I.~{Kravchenko}}, \bibinfo{author}{G.~M. {Frichter}},
  \bibinfo{author}{D.~{Seckel}}, \bibinfo{author}{G.~M. {Spiczak}},
  \bibinfo{author}{J.~{Adams}}, \bibinfo{author}{S.~{Seunarine}},
  \bibinfo{author}{C.~{Allen}}, \bibinfo{author}{A.~{Bean}},
  \bibinfo{author}{D.~{Besson}}, \bibinfo{author}{D.~J. {Box}},
  \bibinfo{author}{R.~{Buniy}}, \bibinfo{author}{J.~{Drees}},
  \bibinfo{author}{D.~{McKay}}, \bibinfo{author}{J.~{Meyers}},
  \bibinfo{author}{L.~{Perry}}, \bibinfo{author}{J.~{Ralston}},
  \bibinfo{author}{S.~{Razzaque}}, \bibinfo{author}{D.~W. {Schmitz}},
\newblock \bibinfo{title}{{Performance and simulation of the RICE detector}},
\newblock \bibinfo{journal}{Astropart. Phys.} \bibinfo{volume}{19}
  (\bibinfo{year}{2003}) \bibinfo{pages}{15--36}.
\bibitem[{Allison et~al.(2016)}]{allison2016performance:2015eky}
\bibinfo{author}{P.~Allison}, et~al.,
\newblock \bibinfo{title}{{Performance of two Askaryan Radio Array stations and
  first results in the search for ultrahigh energy neutrinos}},
\newblock \bibinfo{journal}{Physical Review D} \bibinfo{volume}{93}
  (\bibinfo{year}{2016}) \bibinfo{pages}{082003}.
\bibitem[{Avva et~al.(2015)Avva, Kovac, Miki, Saltzberg, and
  Vieregg}]{avva2015situ}
\bibinfo{author}{J.~Avva}, \bibinfo{author}{J.~M. Kovac},
  \bibinfo{author}{C.~Miki}, \bibinfo{author}{D.~Saltzberg},
  \bibinfo{author}{A.~G. Vieregg},
\newblock \bibinfo{title}{{An in situ measurement of the radio-frequency
  attenuation in ice at Summit Station, Greenland}},
\newblock \bibinfo{journal}{Journal of Glaciology} \bibinfo{volume}{61}
  (\bibinfo{year}{2015}) \bibinfo{pages}{1005--1011}.
\bibitem[{Pearce and Walker(1967)}]{pearce1967empirical}
\bibinfo{author}{D.~Pearce}, \bibinfo{author}{J.~Walker},
\newblock \bibinfo{title}{{An empirical determination of the relative
  dielectric constant of the Greenland Ice Cap}},
\newblock \bibinfo{journal}{Journal of Geophysical Research}
  \bibinfo{volume}{72} (\bibinfo{year}{1967}) \bibinfo{pages}{5743--5747}.
\bibitem[{Mellor and {Cold Regions Research and Engineering Laboratory
  (U.S.)}(1974)}]{mellor1974review}
\bibinfo{author}{M.~Mellor}, \bibinfo{author}{{Cold Regions Research and
  Engineering Laboratory (U.S.)}}, \bibinfo{title}{A Review of Basic Snow
  Mechanics}, \bibinfo{publisher}{U.S. Army Cold Regions Research and
  Engineering Laboratory}, \bibinfo{year}{1974}.
\bibitem[{Gerling et~al.(2017)Gerling, L\"owe, and {van
  Herwijnen}}]{doi:10.1002/2017GL075110}
\bibinfo{author}{B.~Gerling}, \bibinfo{author}{H.~L\"owe},
  \bibinfo{author}{A.~{van Herwijnen}},
\newblock \bibinfo{title}{{Measuring the Elastic Modulus of Snow}},
\newblock \bibinfo{journal}{Geophysical Research Letters} \bibinfo{volume}{44}
  (\bibinfo{year}{2017}) \bibinfo{pages}{11,088--11,096}.
\bibitem[{{Hanson} et~al.(2015){Hanson}, {Barwick}, {Berg}, {Besson}, {Duffin},
  {Klein}, {Kleinfelder}, {Reed}, {Roumi}, {Stezelberger}, {Tatar}, {Walker},
  and {Zou}}]{hansJGlac}
\bibinfo{author}{J.~C. {Hanson}}, \bibinfo{author}{S.~W. {Barwick}},
  \bibinfo{author}{E.~C. {Berg}}, \bibinfo{author}{D.~Z. {Besson}},
  \bibinfo{author}{T.~J. {Duffin}}, \bibinfo{author}{S.~R. {Klein}},
  \bibinfo{author}{S.~A. {Kleinfelder}}, \bibinfo{author}{C.~{Reed}},
  \bibinfo{author}{M.~{Roumi}}, \bibinfo{author}{T.~{Stezelberger}},
  \bibinfo{author}{J.~{Tatar}}, \bibinfo{author}{J.~A. {Walker}},
  \bibinfo{author}{L.~{Zou}},
\newblock \bibinfo{title}{{Radar Absorption, Basal Reflection, Thickness, and
  Polarization Measurements from the Ross Ice Shelf}},
\newblock \bibinfo{journal}{Journal of Glaciology} \bibinfo{volume}{61, 227}
  (\bibinfo{year}{2015}).
\bibitem[{Bogorodsky et~al.(1985)Bogorodsky, Bentley, and Gudmansen}]{Bogo85}
\bibinfo{author}{V.~V. Bogorodsky}, \bibinfo{author}{C.~R. Bentley},
  \bibinfo{author}{P.~E. Gudmansen}, \bibinfo{title}{{Radioglaciology}},
  volume~\bibinfo{volume}{1} of \textit{\bibinfo{series}{1}},
  \bibinfo{publisher}{Reidel}, \bibinfo{address}{P.O. Box 17, 3300 AA
  Dordrecht, Holland}, \bibinfo{edition}{1} edition, \bibinfo{year}{1985}.
\bibitem[{Gerland et~al.(1999)Gerland, Oerter, Kipfstuhl, Wilhelms, Miller, and
  Miners}]{gerland_oerter_kipfstuhl_wilhelms_miller_miners_1999}
\bibinfo{author}{S.~Gerland}, \bibinfo{author}{H.~Oerter},
  \bibinfo{author}{J.~Kipfstuhl}, \bibinfo{author}{F.~Wilhelms},
  \bibinfo{author}{H.~Miller}, \bibinfo{author}{W.~D. Miners},
\newblock \bibinfo{title}{{Density log of a 181 m long ice core from Berkner
  Island, Antarctica}},
\newblock \bibinfo{journal}{Annals of Glaciology} \bibinfo{volume}{29}
  (\bibinfo{year}{1999}) \bibinfo{pages}{215–219}.
\bibitem[{Dowdeswell and Evans(2004)}]{0034-4885-67-10-R03}
\bibinfo{author}{J.~A. Dowdeswell}, \bibinfo{author}{S.~Evans},
\newblock \bibinfo{title}{Investigations of the form and flow of ice sheets and
  glaciers using radio-echo sounding},
\newblock \bibinfo{journal}{Reports on Progress in Physics}
  \bibinfo{volume}{67} (\bibinfo{year}{2004}) \bibinfo{pages}{1821}.
\bibitem[{Maeno and Ebinuma(1983)}]{maeno1983pressure}
\bibinfo{author}{N.~Maeno}, \bibinfo{author}{T.~Ebinuma},
\newblock \bibinfo{title}{{Pressure sintering of ice and its implication to the
  densification of snow at polar glaciers and ice sheets}},
\newblock \bibinfo{journal}{The Journal of Physical Chemistry}
  \bibinfo{volume}{87} (\bibinfo{year}{1983}) \bibinfo{pages}{4103--4110}.
\bibitem[{Kravchenko et~al.(2004)Kravchenko, Besson, and
  Meyers}]{kravchenko2004situ}
\bibinfo{author}{I.~Kravchenko}, \bibinfo{author}{D.~Besson},
  \bibinfo{author}{J.~Meyers},
\newblock \bibinfo{title}{{In situ index-of-refraction measurements of the
  South Polar firn with the RICE detector}},
\newblock \bibinfo{journal}{Journal of Glaciology} \bibinfo{volume}{50}
  (\bibinfo{year}{2004}) \bibinfo{pages}{522--532}.
\bibitem[{Gerhardt et~al.(2010)Gerhardt, Klein, Stezelberger, Barwick,
  Dookayka, Hanson, and Nichol}]{Gerhardt:2010js}
\bibinfo{author}{L.~Gerhardt}, \bibinfo{author}{S.~Klein},
  \bibinfo{author}{T.~Stezelberger}, \bibinfo{author}{S.~Barwick},
  \bibinfo{author}{K.~Dookayka}, \bibinfo{author}{J.~Hanson},
  \bibinfo{author}{R.~Nichol},
\newblock \bibinfo{title}{{A prototype station for ARIANNA: a detector for
  cosmic neutrinos}},
\newblock \bibinfo{journal}{Nucl. Instrum. Meth.} \bibinfo{volume}{A624}
  (\bibinfo{year}{2010}) \bibinfo{pages}{85--91}.
\bibitem[{Abdul et~al.(2017)}]{abdul2017measurement}
\bibinfo{author}{U.~Abdul}, et~al.,
\newblock \bibinfo{title}{{Measurement of the real dielectric permittivity
  epsilon\_r of glacial ice}},
\newblock \bibinfo{journal}{arXiv preprint arXiv:1712.03301}
  (\bibinfo{year}{2017}).
\bibitem[{Barwick et~al.(2015)}]{Barwick:2014rca}
\bibinfo{author}{S.~W. Barwick}, et~al.,
\newblock \bibinfo{title}{{Design and Performance of the ARIANNA Hexagonal
  Radio Array Systems}},
\newblock \bibinfo{journal}{IEEE Trans. Nucl. Sci.} \bibinfo{volume}{62}
  (\bibinfo{year}{2015}) \bibinfo{pages}{2202--2215}.
\bibitem[{Neal(1979)}]{neal_1979}
\bibinfo{author}{C.~S. Neal},
\newblock \bibinfo{title}{{The Dynamics of the Ross Ice Shelf Revealed by Radio
  Echo-Sounding}},
\newblock \bibinfo{journal}{Journal of Glaciology} \bibinfo{volume}{24}
  (\bibinfo{year}{1979}) \bibinfo{pages}{295–307}.
\bibitem[{Barwick et~al.(2017)}]{barwick2017radio}
\bibinfo{author}{S.~W. Barwick}, et~al.,
\newblock \bibinfo{title}{{Radio detection of air showers with the ARIANNA
  experiment on the Ross Ice Shelf}},
\newblock \bibinfo{journal}{Astroparticle Physics} \bibinfo{volume}{90}
  (\bibinfo{year}{2017}) \bibinfo{pages}{50--68}.
\bibitem[{Hanson(2013)}]{JordanHansonUCIphdthesis}
\bibinfo{author}{J.~Hanson}, \bibinfo{title}{{The Performance and Initial
  Results of the ARIANNA Prototype}}, Ph.D. thesis, University of California,
  Irvine, \bibinfo{year}{2013}.
\bibitem[{Barwick et~al.(2015)}]{Barwick:2014boa}
\bibinfo{author}{S.~W. Barwick}, et~al.,
\newblock \bibinfo{title}{{Time Domain Response of the ARIANNA Detector}},
\newblock \bibinfo{journal}{Astropart. Phys.} \bibinfo{volume}{62}
  (\bibinfo{year}{2015}) \bibinfo{pages}{139--151}.
\bibitem[{Allison et~al.(2012)}]{allison2012design}
\bibinfo{author}{P.~Allison}, et~al.,
\newblock \bibinfo{title}{{Design and initial performance of the Askaryan Radio
  Array prototype EeV neutrino detector at the South Pole}},
\newblock \bibinfo{journal}{Astroparticle Physics} \bibinfo{volume}{35}
  (\bibinfo{year}{2012}) \bibinfo{pages}{457--477}.
\bibitem[{Gorham et~al.(2017)Gorham, Allison, Banerjee, Batten, Beatty, Belov,
  Besson, Binns, Bugaev, Cao et~al.}]{gorham2017hical}
\bibinfo{author}{P.~Gorham}, \bibinfo{author}{P.~Allison},
  \bibinfo{author}{O.~Banerjee}, \bibinfo{author}{L.~Batten},
  \bibinfo{author}{J.~Beatty}, \bibinfo{author}{K.~Belov},
  \bibinfo{author}{D.~Besson}, \bibinfo{author}{W.~Binns},
  \bibinfo{author}{V.~Bugaev}, \bibinfo{author}{P.~Cao}, et~al.,
\newblock \bibinfo{title}{{The HiCal 2 Instrument: Calibration and Antarctic
  Surface Reflectivity Measurement for the ANITA Experiment}},
\newblock \bibinfo{journal}{arXiv preprint arXiv:1710.11175}
  (\bibinfo{year}{2017}).
\bibitem[{Casey et~al.(2014)Casey, Fudge, Neumann, Steig, Cavitte, and
  Blankenship}]{casey20141500}
\bibinfo{author}{K.~A. Casey}, \bibinfo{author}{T.~Fudge},
  \bibinfo{author}{T.~Neumann}, \bibinfo{author}{E.~Steig},
  \bibinfo{author}{M.~Cavitte}, \bibinfo{author}{D.~Blankenship},
\newblock \bibinfo{title}{{The 1500 m South Pole ice core: recovering a 40 ka
  environmental record}},
\newblock \bibinfo{journal}{Annals of Glaciology} \bibinfo{volume}{55}
  (\bibinfo{year}{2014}) \bibinfo{pages}{137--146}.

\end{thebibliography}
\end{document}